\documentclass[11pt,titlepage]{article} 
\usepackage[a4paper]{geometry}
\usepackage{graphicx}
\usepackage{amssymb,amsmath} 

\usepackage{color,ulem}

\title{
Experimental free energy measurements of kinetic molecular states using fluctuation theorems
} 
\author{$^1$Anna Alemany, $^2$Alessandro Mossa, $^3$Ivan Junier, $^{1,4}$Felix Ritort} 
\date{$^1$Small Biosystems Lab, Departament de F\'{\i}sica Fonamental, 
Universitat de Barcelona,  Avda. Diagonal 647, 08028 Barcelona, Spain\\
$^2$Department of Physics and Astronomy, University of Aarhus,
 Aarhus C, Denmark \\
$^3$Centre de Regulaci\'o Gen\`omica (CRG), C/ Dr. Aiguader 88, 08003 Barcelona, Spain \\
$^4$CIBER-BBN de Bioingenier\'ia, Biomateriales y Nanomedicina, 
Instituto de Salud Carlos III, Madrid, Spain\\
}

\begin{document}

\maketitle

{\bf Recent advances in non-equilibrium statistical mechanics and single molecule technologies make it possible to extract free energy differences from  irreversible work measurements in pulling experiments. To date, free energy recovery has been focused on native or equilibrium molecular states, whereas free energy measurements of kinetic states (i.e. finite lifetime states that are generated dynamically and are metastable) have remained unexplored. 
Kinetic states can play an important role in various domains of physics, such as nanotechnology or condensed matter physics. In biophysics, there are many examples where they determine the fate of molecular reactions: 
protein and peptide-nucleic acid binding, specific cation binding, antigen-antibody interactions,
transient states in enzymatic reactions
or the formation of transient intermediates and non-native structures in
molecular folders. 
Here we demonstrate that it is possible to obtain free energies of kinetic states
by applying extended fluctuation relations.
This is shown by using optical tweezers to mechanically unfold and refold DNA structures exhibiting 
intermediate and misfolded kinetic states.
}

Kinetic states are observed under non-equilibrium conditions and have higher free energies than native states. Yet, they can be crucial, as shown by the role that misfolded proteins play in numerous severe diseases \cite{Dobson}. 
The measurement of the free energy of formation of kinetic states is therefore a central question in
biophysics. Recent theoretical developments known as fluctuation relations \cite{Jarzynski1997,Crooks2000,Ritort2006,Woodside2008,Jarzynski2011} have been applied to extract free energy differences of equilibrium states from irreversible work measurements. Applications include the measurement of the free energy of formation of RNA and DNA hairpins \cite{CollinRitort2005}; the determination of the stability of native domains in proteins \cite{ShankCecconi2010}; the measurement of mechanical torque in rotary motors \cite{Hayashi2010}; the conversion of information into work in systems under feedback control \cite{Toyabe2010}; or the recovery of free energy landscapes from unidirectional work measurements \cite{HummerSzabo2000, Woodside2011}. 

The characterization of kinetic states under non-equilibrium conditions remains a challenging problem.
Here we use a recently introduced extended fluctuation relation (EFR) to extract free energies of kinetic
states and thermodynamic branches using irreversible work measurements \cite{Maragakis, Junier}.
In the EFR, a kinetic state is a \textit{partially equilibrated} region of configurational space, meaning that during a finite timescale the system is confined and thermalized within that region \cite{Palmer}. This is mathematically described by a Boltzmann--Gibbs distribution restricted to configurations contained in that region (Fig. 1a). 

Let $A,B$ denote any two kinetic states and $\lambda$ a control parameter.
We consider a forward (F) non-equilibrium process, where the system starts in partial equilibrium in $A$ at $\lambda_0$, and its time-reversed (R), where the partial equilibrium condition is required over $B$ at $\lambda_1$.
In the F process
$\lambda$ varies from $\lambda_0$ to $\lambda_1$ during a time $\tau$
according to a predetermined protocol $\lambda(t)$. For the R process the time-reversed protocol $\lambda(\tau-t)$ is used. The EFR reads \cite{Junier}: 
\begin{equation}
\frac{\phi_F^{A\rightarrow B}}{\phi_R^{A\leftarrow B}}\frac{P_F^{A \rightarrow B}(W)}{P_R^{A\leftarrow B}(-W)}
=\exp\left[\frac{W-\Delta G_{AB}}{k_{\rm B}T}\right]
\label{eq: EFR}
\end{equation}
where $\Delta G_{AB}=G_B(\lambda_1)-G_A(\lambda_0)$ is the free energy difference between kinetic states $B$ at $\lambda_1$ and $A$ at $\lambda_0$; $P_F^{A\rightarrow B}(W)$ ($P_R^{A\leftarrow B}(-W)$)
denotes the partial work distribution for the F (R) process over
the fraction of paths $\phi_F^{A\rightarrow B}$ ($\phi_R^{A\leftarrow B}$) starting in $A$
($B$) at $\lambda_0$ ($\lambda_1$) and ending in $B$ ($A$) at $\lambda_1$ ($\lambda_0$); $k_{\rm B}$ is the Boltzmann constant and
$T$ the temperature of the environment.

We applied equation \eqref{eq: EFR} to extract free energy differences of kinetic states from mechanical unfolding/folding
experiments performed on DNA hairpins, which are model systems easy to design and synthesize.
Their free energies of formation
can be predicted using the nearest-neighbor (NN) model with the unified-oligonucleotide (UO) set of parameters \cite{Santalucia,Zuker2003} or with recently derived energies from unzipping experiments \cite{HugBizForSmiBusRit10} (Methods, Supplementary Section S1).
Molecules exhibiting two types of kinetic states were investigated (Fig. 1b):
molecules I1 and I2 have intermediate kinetic states on-pathway to the native state, and molecules M1 and M2 have misfolded kinetic states off-pathway to the native state. 
To establish the validity of our approach, we first show results for
molecules I1 and M1 where free energies measured from the EFR applied
to non-equilibrium pulling experiments can be compared with
free energies obtained from equilibrium hopping experiments. The method is then
applied to molecules I2 and M2 where irreversibility or low
signal-to-noise ratio in hopping traces preclude equilibrium based
free energy measurements.

The experimental setup is shown in Fig. 1c \cite{HugBizForSmiBusRit10,Forns2010} (Methods).
We steer up and down the position of
the optical trap to mechanically unfold and refold
the DNA hairpin, and measure the force acting on the
hairpin as a function of the relative trap-pipette distance, which
is the control parameter $\lambda$ \cite{MosLorHugRit09}.
We measure the work as the area below the force-distance curve (hereafter
referred as FDC, inset of Fig. 2a) along many trajectories.
Throughout this paper unfolding (folding) corresponds to the F (R) process.

First we apply equation \eqref{eq: EFR}
to hairpin I1 characterized by three conformational states (Fig. 1b, Supplementary Section S1):
native (N), intermediate (I) and unfolded (U). 
Experimental hopping traces measured under equilibrium conditions and non-equilibrium FDCs exhibit three force branches corresponding to the three states (Fig. 2a, Supplementary Video VI1 and Sections S2, S3). 
Fig. 2b shows the partial work distributions 
measured from a collection of FDCs by taking $\lambda_0$=0, where $A$=N, and $\lambda_1$=55.6 nm, where the three states are observed ($B$=N, I or U). These partial work distributions satisfy equation \eqref{eq: EFR}
(Supplementary Section S4).
Hysteresis effects are stronger for $B$=I, U than for $B$=N, 
as the timescale related to the pulling protocol is typically shorter than the timescale for crossing the kinetic barrier separating two states. 
The acceptance ratio method \cite{Bennett,ShiBaiHooPan03} applied to extract the free energy differences between states gives $\Delta G_{\rm NU}\simeq\Delta G_{\rm NN}$, and $\Delta G_{\rm NI}$ lies 2 $k_{\rm B}T$ above (Fig. 2c, Methods).
Fig. 2d shows the reconstruction of the three
thermodynamic branches by fixing $\lambda_0$=0 and varying $\lambda_1$ between 45 and 65 nm. 
The vertical dashed-dotted line at $\lambda_c$=55.6 nm indicates the coexistence point of N and U. The full equilibrium free energy of the system, defined as $\Delta G=-k_{\rm B}T\log\left(e^{-\Delta G_{\rm NN}/k_{\rm B}T}+e^{-\Delta G_{\rm NI}/k_{\rm B}T}+e^{-\Delta G_{\rm NU}/k_{\rm B}T}\right)$, has also been measured (Fig. 2d, black line). The right inset in Fig. 2d shows the free energy of each state measured relative to $\Delta G$. 
For $\lambda<\lambda_c$ ($\lambda>\lambda_c$) N (U) is the most stable state, while I is never the absolute free energy minimum for any $\lambda$. 
The left inset in Fig. 2d shows the contribution of $\phi_F^{A\rightarrow B}/\phi_R^{A\leftarrow B}$ to the measured free energies throughout the $\lambda$ range.
Dropping this term or misidentifying states along the FDC leads to wrong free energy predictions (Supplementary Sections S5, S6). 

By subtracting the elastic contributions due to stretching the handles and the released ssDNA
(Methods, Supplementary Section S7) we extract the free energies of formation of the
different structures with respect to the random coil state at zero force. We get $\Delta G_{\rm NU}^0=55\pm3$ $k_{\rm B}T$  and $\Delta G_{\rm NI}^0=30\pm3$ $k_{\rm B}T$, in agreement with free energy predictions and results from equilibrium-based hopping experiments (Table 1).

Next we study hairpin M1, which can fold into two
unrelated structures (Fig. 1b, Supplementary Section S1): the native (N) and the misfolded (M). 
Equilibrium hopping experiments exhibit very fast kinetics and two clearly separated hopping regions: in one region N coexists with an intermediate state on-pathway; in the other region M and U coexist with another intermediate (Supplementary Section 2). For simplicity reasons, we chose not to characterize these intermediate states.
In non-equilibrium experiments, two FDC patterns are identified corresponding to the two structures (Fig. 3a, Supplementary Video VM1 and Section S3). 
In contrast to the unfolding/folding cycles that start and end in N, those that start and end in M show almost no hysteresis (Fig. 3b), indicating low kinetic barriers between M and U. Due to kinetic competition of loop formation between M and N, M has a basin of attraction larger than N during folding ($\simeq$80\% of folding trajectories end in M), lower thermodynamic stability and larger molecular extension at low forces (Fig. 2a, inset).

In Fig. 3c we apply the acceptance ratio method to recover the free energy differences between U at $\lambda_1$=130 nm and M (lower set of measurements) or N (upper set) at $\lambda_0$=0 nm. 
By subtracting the handles and ssDNA contributions we extract
the free energy of formation of each structure at zero force, obtaining $\Delta G_{\rm MU}^0=47\pm2$ $k_{\rm B}T $ and $\Delta G_{\rm NU}^0=62\pm3$ $k_{\rm B}T$. The distribution of free energies for different molecules and pulling speeds is shown in Fig. 3d. The difference between the average of both distributions is in agreement with predictions based on the NN model and with results from equilibrium-based hopping experiments (Table 1). 

Finally, to illustrate the power of the
method we show the case of molecules I2 and M2, where it is not possible to recover free energies using equilibrium based methods (Supplementary Section S3).

Hairpin I2 has two intermediate states on pathway, hereafter referred as I' and I'' \cite{Engel2011} (Supplementary Section S1). In pulling experiments four force branches, corresponding to states N, I', I'' and U are distinguished (Fig. 4a, Supplementary Video VI2 and Sections S3, S4). 
In order to measure forward and reversed partial work distributions for the four states we pull back and forth the molecule between $\lambda_0$=0 (where the molecule is in equilibrium at N) and $\lambda_1$=183 nm
(where the molecule is partially equilibrated at states N, I', I'', U; Fig. 4a,b). 
This protocol is subtly different from the standard pulling experiments we did for the rest of molecules (I1, M1, M2) where the molecule is never in an intermediate state at initial and final values of $\lambda$. Due to the larger hysteresis exhibited by this molecule (Supplementary Section S7), the standard protocol does not generate reverse trajectories that sample all four states for any value of $\lambda$.

In table 1 we show the values of the free energy of formation of the different kinetic states
obtained with the EFR. The size of the error bars is comparable to the discrepancy between the
free energy predictions using the NN model with the UO set of parameters \cite{Santalucia,Zuker2003} and unzipping data
\cite{HugBizForSmiBusRit10}. 
To evaluate the free energy branches of the different states (Fig. 4c) we could repeat the experiment
for different final values of $\lambda$ and measure the corresponding $\Delta G_{{\rm N}B}(\lambda)$, $B$=N, I', I'' or U. For simplicity we use an extended version of the Jarzynski equality (EJE) obtained by multiplying the EFR with the reversed work distribution and integrating over the work (Supplementary Section S8),
\begin{equation}
\Delta G_{AB}=-k_{\rm B}T\log\left(\frac{\phi_F^{A\rightarrow B}}{\phi_R^{A\leftarrow B}}\right)-k_{\rm B}T\log \left\langle e^{-\frac{W}{k_{\rm B}T}}\right\rangle_F^{A\rightarrow B}\label{eq: EJE}
\end{equation}
Equation \eqref{eq: EJE} only requires data from the F process and we apply it to pulling experiments recorded by setting extreme values of $\lambda$ (light curves in Fig. 4a).
Similarly to the Jarzynski estimator \cite{Jarzynski1997}, the EJE is strongly biased \cite{PalassiniRitort2011}. To estimate the magnitude of the bias we took the difference between the free energy $\Delta G_{{\rm N}B}$ obtained using equation \eqref{eq: EJE} to the one obtained with the acceptance ratio method in pulling experiments where kinetic states are partially equilibrated at $\lambda_1$ (dark curves in Fig. 4a). Therefore, from the free energy branches obtained using equation \eqref{eq: EJE} we subtracted this estimated bias for each state (we assumed it to be equal for all values of $\lambda_1$, Supplementary Section S8).
In contrast to I1, kinetic intermediates found in I2 become the most stable states in a given range of $\lambda$. For low values of $\lambda$, stability is determined by N, and as $\lambda$ increases stability shifts to I', I'' and finally to U (Fig. 4c).

Hairpin M2 can fold
into one native structure (N) and two misfolded structures (M', M'')
following alternative folding pathways (Fig. 5a, Supplementary Section S1).
Whereas it is easy to identify trajectories that fold into N (red/blue FDCs in Fig. 5a left, $\sim$50\% of trajectories), distinguishing trajectories that misfold into M' or M'' is not straightforward. Careful inspection reveals two different patterns of unfolding curves that start at a misfolded state: either the molecule unfolds quasi-reversibly without intermediates (purple FDC in Fig. 5a middle, $\sim$30\% of trajectories), or it folds back to N before it unfolds (cyan FDC in Fig. 5a right, $\sim$20\% of trajectories) \cite{LiBusTin07}. 
We interpret the former as trajectories following the M'$\rightarrow$U pathway and the latter following the M''$\rightarrow$N$\rightarrow$U pathway (Supplementary Video VM2). 
This is supported by two facts (Supplementary Section S5). First, M' consists of four small hairpins that confer low mechanical stability to the structure that gently unfolds under tension. Second, M'' has a large stem in common with N (Fig. 5a, top) which is surrounded by two small hairpins with low mechanical stability.
Once these two hairpins unfold around 9-10 pN, force remains low enough for the molecule to fold back to N before unfolding.
Combining equation \eqref{eq: EFR}, the partial work distributions (Fig. 5b) and handles and ssDNA  elastic contributions leads to the free energy values 
$\Delta G_{\rm NU}^0=94\pm2$, $\Delta G_{\rm M'U}^0=60\pm3$ and $\Delta G_{\rm M''U}^0=70\pm3$ $k_{\rm B}T$, in agreement with theoretical predictions (Table 1).
Fig. 5c shows the reconstruction of the four thermodynamic branches relative to the full equilibrium free energy of the system, $\Delta G=-k_{\rm B}T\log\sum_{A=\rm{N,M',M'',U}} e^{-\Delta G_{A\rm{U}}/k_{\rm B}T}$, by fixing $\lambda_1$=230 nm and varying $\lambda_0$ between 0 and 150 nm.

Summarizing, we have shown how the EFR can be used to extract
free energies of non-equilibrium kinetic structures in DNA hairpins exhibiting
intermediate and misfolded states. 
The method accurately works in far from equilibrium situations and when equilibrium experiments are insufficient to characterize non-native states.
There are two main differences between the EFR in equation \eqref{eq: EFR} and the Crooks relation \cite{Crooks2000}:
the partial work distributions and the prefactor $\phi_F^{A\rightarrow B}/\phi_R^{A\leftarrow B}$,
which introduces the additional correction $-k_{\rm B}T\log(\phi_F^{A\rightarrow B}/\phi_R^{A\leftarrow B})$
into the Crooks estimation of the free energy difference between kinetic states.
The omission of such correction yields wrong relative thermodynamic stabilities for the free energy branches of the 
different kinetic states (Supplementary Section S7) \cite{Junier}.
Moreover, for the case of misfolded structures that apparently unfold/misfold
reversibly (Fig. 3a right and Fig. 5a middle), $\Delta G_{{\rm MU}}$ is not just equal to the
measured reversible work during unfolding since since the term $k_{\rm B}T\log \phi_R^{{\rm M}\leftarrow {\rm U}}$ must be added (here $\phi_F^{{\rm M}\rightarrow {\rm U}}=1$ as F processes always end at U). Although this
correction is small for states M (M1) and M'' (M2) ($\sim$ 0.2 $k_{\rm B}T$ and
1.2 $k_{\rm B}T$ respectively), it is important in situations where $\phi_R^{{\rm M}\leftarrow {\rm U}}\ll1$, even if very low hysteresis is obtained between the F and R processes. For example, the neglection of a 1\% misfolding probability would underestimate by 4.5 $k_{\rm B}T$ the free energy of formation of the misfolded state.

The main limitation of the method is the identification of kinetic states from the measured signal. 
In this regard, a combination of fluorescence techniques, such as FRET, with force measurements, and the application of advanced statistical methods (e.g. hidden Markov models or Bayesian inference) might be very useful.
Our methodology should find many applications that range from molecular biophysics to condensed matter physics. Any situation where equilibrium experiments are unpractical should be treatable with different versions of equation (1). To start with, 
the method can be employed for measuring free energies of kinetic structures that appear in many molecular reactions, such as RNA, proteins, and many kinetic states related to intermolecular binding, or transient non-equilibrium
states that are essential in polymerization reactions (e.g. ATP or ADP bound states in motor proteins).

\section*{Methods}

\subsection*{Molecular synthesis}

The designed DNA molecules linked to 29 bp dsDNA-handles were synthesized as described in \cite{Forns2010}.
For the specific attachments to the DNA molecular construction we used streptavidin-coated polystyrene microspheres (1.87 $\mu$m, Spherotech, Libertyville, IL) and protein G microspheres (3.0-3.4 $\mu$m; G. Kisker Gbr, Products for Biotechnologie, Steinfurt, Germany) coated with anti-digoxigenin polyclonal antibodies (Roche Applied Science, Spain).
Attachment to the anti-digoxigenin microspheres was achieved first by incubating the beads
with the tether DNA. The second attachment was achieved in the fluidics chamber and was
accomplished by bringing a trapped anti-digoxigenin and an immobilized streptavidin microsphere close to
each other.

\subsection*{Bennett acceptance ratio method}

This method is used to estimate the free energy difference $\Delta G_{AB}$ between two states from non-equilibrium work measurements. Given a set of $n_F$($n_R$) forward(reversed) work measurements, it is shown in \cite{Bennett,ShiBaiHooPan03} that the solution of the following transcendental equation:
\begin{align}
\label{eq: bennet}
\beta u&=z(u) \nonumber\\
&=-\log\left(\frac{\phi_F^{A\rightarrow B}}{\phi_R^{A\leftarrow B}}\right)+z_R(u)-z_F(u)
\end{align}
where
\begin{subequations}
\begin{equation}
z_R(u)=\log\frac{1}{n_R}\sum_{i=1}^{n_R}\left( \frac{e^{-\beta W_i}}{1+\frac{n_F}{n_R}e^{-\beta(W_i+u)}}\right)
\end{equation} 
\begin{equation}
z_F(u)=\log\frac{1}{n_F}\sum_{i=1}^{n_F}\left( \frac{1}{1+\frac{n_F}{n_R}e^{\beta(W_i-u)}} \right)
\end{equation}
\end{subequations}
minimizes the statistical variance of the free energy estimation for $u=\Delta G_{AB}$. 

The right-hand side of equation (\ref{eq: bennet}) is expected to provide a constant function near the solution of the transcendental equation, as shown in Figs. 2c and 3c for each branch. 

\subsection*{Free energy recovery at zero force}

The solution of the Bennett acceptance ratio method gives the free energy difference between state $A$ at $\lambda_0$ and state $B$ at $\lambda_1$. In order to recover the free energy at zero force of each structure with respect to the random coil state, $\Delta G_0$, we need to subtract the free energy of stretching the ssDNA, $W_{\rm ssDNA}$, the free energy of orientation of the hairpins stem, $W_{\rm stem}$, and the reversible work performed to stretch the handles and displace the bead in the optical trap, $W_{\rm hb}$:
\begin{equation}
\Delta G_{AB}^0=\Delta G_{AB}-W_{\rm ssDNA}-W_{\rm stem}-W_{\rm hb}
\end{equation}
To compute the work needed to reversibly stretch the ssDNA, $W_{\rm ssDNA}=\int fdx$, we use the non-extensible worm-like chain elastic model with a persistence length equal to 1.3$\pm$0.2 nm and a contour length equal to the contour length of the molecule. The free energy of the stem orientation is evaluated using the freely-jointed chain with a Kuhn length equal to the diameter of the hairpin at zero force (Supplementary Section S1) \cite{Forns2010}. The short length of the handles allows us to estimate the free energy of the handles and the bead by integrating a linear FDC along the folded branch from the minimum force at $\lambda_0$ to the maximum force at $\lambda_1$, that is, $W_{\rm hb}=(f_{\rm max}^2-f_{\rm min}^2)/2k_{\mathrm{eff}}$, where $k_{\mathrm{eff}}$ is the slope of the FDC \cite{MossaJSTAT2010}. 

\subsection*{Free energy prediction}

In order to obtain the most stable structure of the DNA molecules under study we use the mfold web server \cite{Zuker2003}.
To predict the free energy of formation of each structure we use the nearest-neighbor model (Supplementary Section S1). The base pairing free energies have been derived in thermal denaturation experiments \cite{Santalucia} and independently verified in single molecule experiments \cite{HugBizForSmiBusRit10}.

\hspace{0.5cm}

\textbf{Acknowledgements.}  
A. A. is supported by grant AP2007-00995 (Spanish Research Council). 
A. M. acknowledges funding from Lundbeckfonden.
I. J. is supported by a Novartis grant (CRG).
F. R. is supported by grants FIS2010-19342, Icrea Academia 2008, and Human Frontier Science Program (HFSP, RGP55-2008). 
We thank J. Horowitz and M. Palassini for a careful reading of the manuscript.

\hspace{0.5cm}

\textbf{Author contributions.}
I. J and F. R. designed the experiment. 
A. A. made the measurements.
A. A. and A. M. analyzed the data.
A. A., A. M., I. J. and F. R. wrote the paper.

\hspace{0.5cm}

\textbf{Additional information}
The authors declare no competing financial interests.
Supplementary information accompanies this paper on www.na\-tu\-re.com/na\-tu\-re\-physics. 
Reprints and permissions information is available on-line at www.na\-ture.com/re\-prints.
Correspondence and requests for materials should be addressed to F. R. 

\clearpage

\section*{Fig. Legends}

\subsubsection*{Fig. 1.}

\includegraphics[scale=0.8]{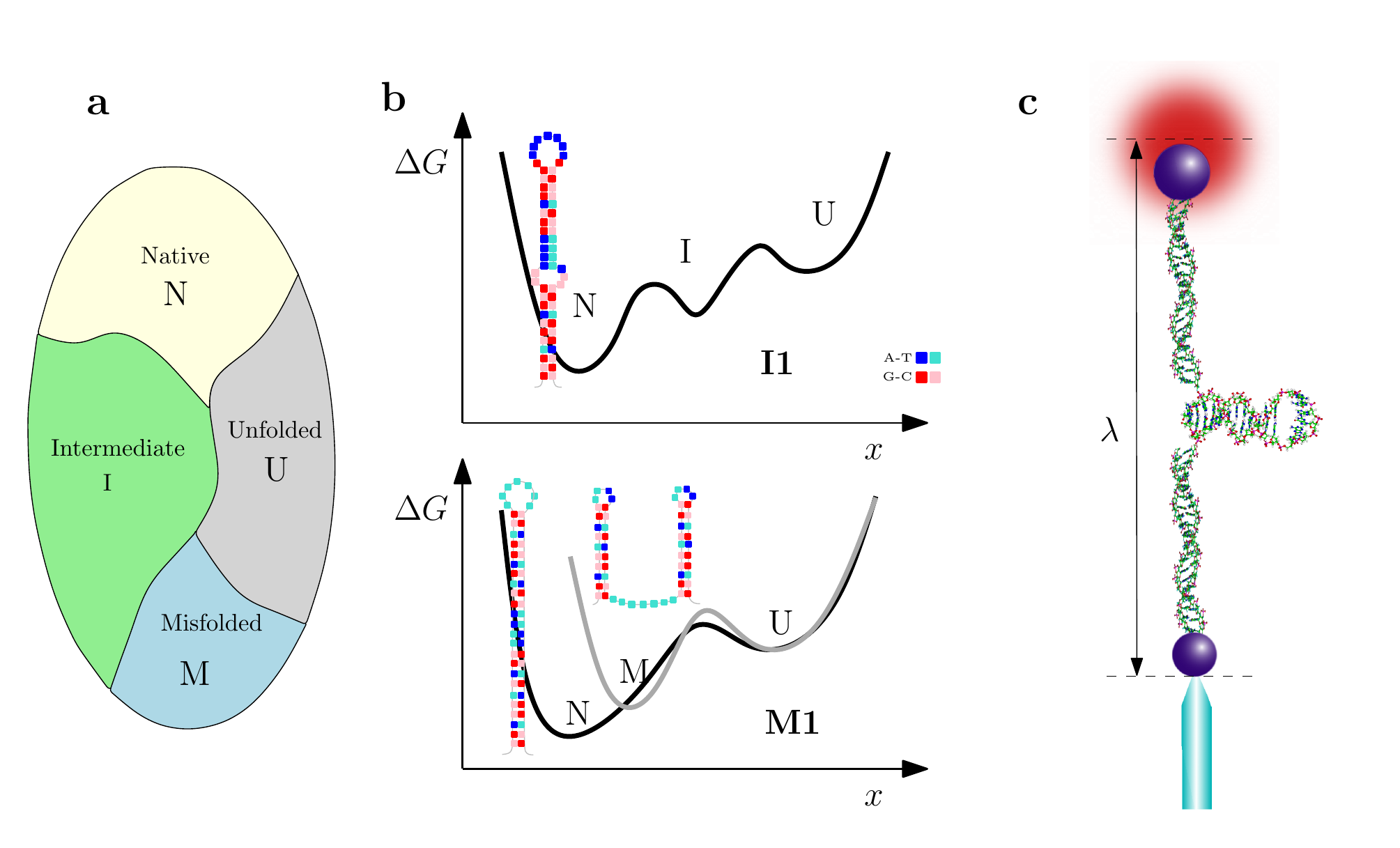}

\textbf{Schematic illustration of the configurational space, molecular free energy landscapes and experimental setup.}
\textbf{a)} Configurational space partitioned into regions that correspond to different molecular kinetic states.
Inside a partially equilibrated region (N, I, U, or M) configurations are sampled according to the Boltzmann--Gibbs distribution; in contrast, the statistical weights of the regions do not necessarily follow an equilibrium distribution.
\textbf{b)} Schematic free energy landscapes for DNA sequences exhibiting an intermediate kinetic state on-pathway (I1, top) and a misfolded kinetic state off-pathway (M1, bottom). For M1, the free energy landscapes of the native (black) and the misfolded (gray) structures are sketched. Free energies and extensions are shown in arbitrary units. 
\textbf{c)} Experimental setup (not to scale). One bead is immobilized in a micropipette by air suction, while the other is captured in an optical trap produced by a highly stable dual-beam optical tweezers apparatus \cite{HugBizForSmiBusRit10}. 

\subsubsection*{Fig. 2.} 

\includegraphics[scale=1]{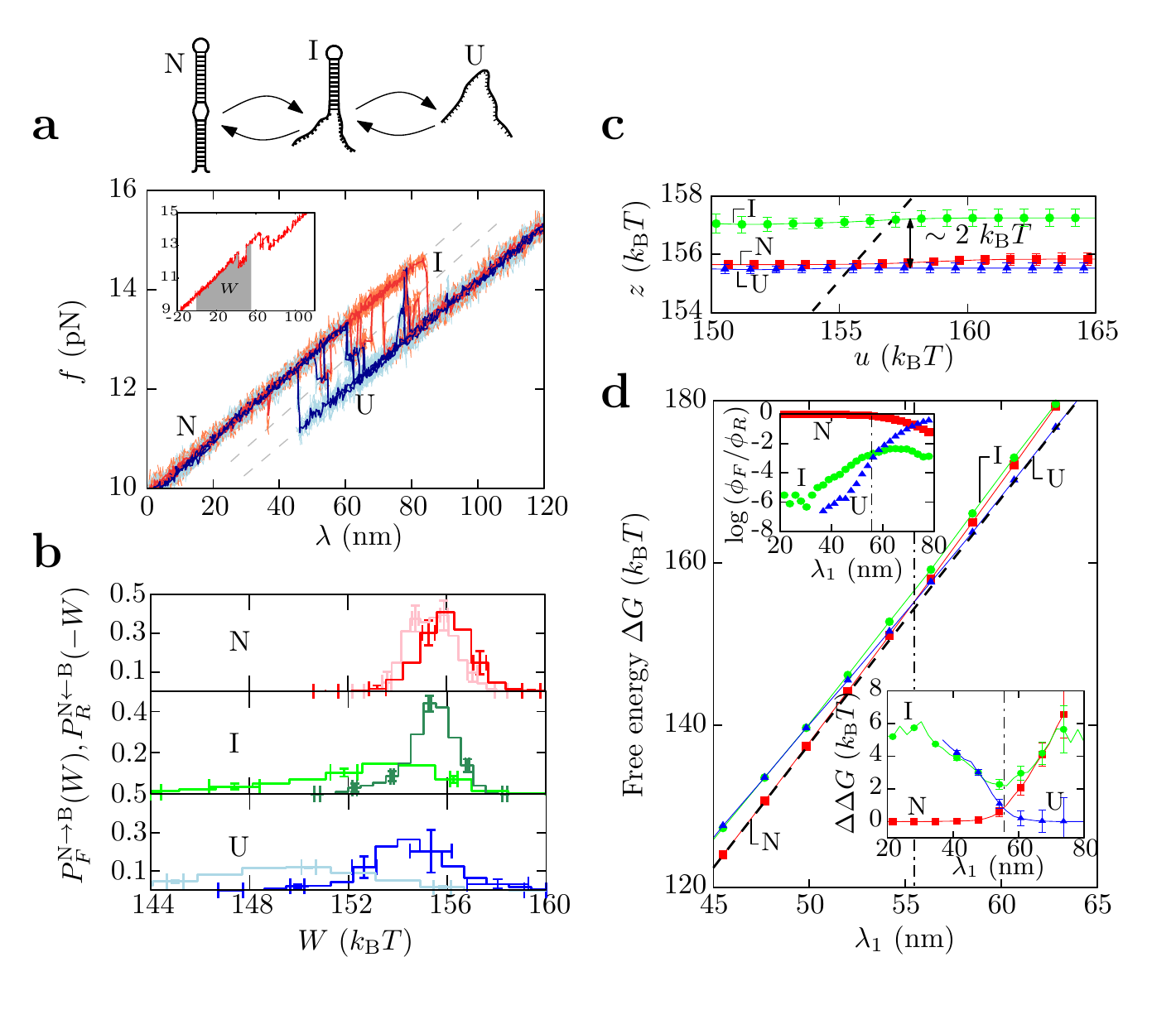}

\textbf{Hairpin I1, with an intermediate state.}
\textbf{a)} Schematic illustration of the mechanical unfolding/folding pathway (top) and collection of unfolding (red) and folding (blue) FDCs recorded at 60 nm/s exhibiting three branches of force corresponding to three states: N, I and U. 
Twelve molecules were pulled at 60 nm/s; between 80 and 385 cycles were collected for each molecule. 
\textit{Inset:} The gray area indicates the work measured for a given FDC between $\lambda_0$=0 and $\lambda_1$=55.6 nm.
\textbf{b)} Partial work histograms for work values measured between $\lambda_0$=0, where $A$=N, and $\lambda_1$=55.6 nm, where $B$=N (red, top panel), $B$=I (green, middle panel) or $B$=U (blue, bottom panel). Dark colors refer to unfolding work distributions and light colors to folding work distributions.
\textbf{c)} The acceptance ratio method (Methods) applied to the work measurements shown in panel b to obtain the free energy differences $\Delta G_{{\rm N}B}$ for $B$=N (red squares), I (green circles) and U (blue triangles).
\textbf{d)} Reconstruction of the free energy branches for states N, I and U (color code as in c) obtained by fixing $\lambda_0$ and letting $\lambda_1$ change between 45 and 65 nm. The black-dashed curve is the full free energy of the system, $\Delta G=-k_{\rm B}T\log\sum_{B={\rm N, I, U}}e^{-\Delta G_{{\rm N}B}/k_{\rm B}T}$.
\textit{Left inset}: Contribution of the prefactor $\log(\phi_F^{{\rm N}\rightarrow B}/\phi_R^{{\rm N}\leftarrow B})$ to the free energy of each kinetic state as a function of $\lambda$. 
\textit{Right inset}: Difference between the free energy of each state, $\Delta G_{{\rm N} B}$, and the full equilibrium free energy, $\Delta G$. Error bars in panels b, c, d indicate the standard statistical deviation computed over 385 cycles for a given molecule. These were obtained using the bootstrap method.

\subsubsection*{Fig. 3.}

\includegraphics[scale=1]{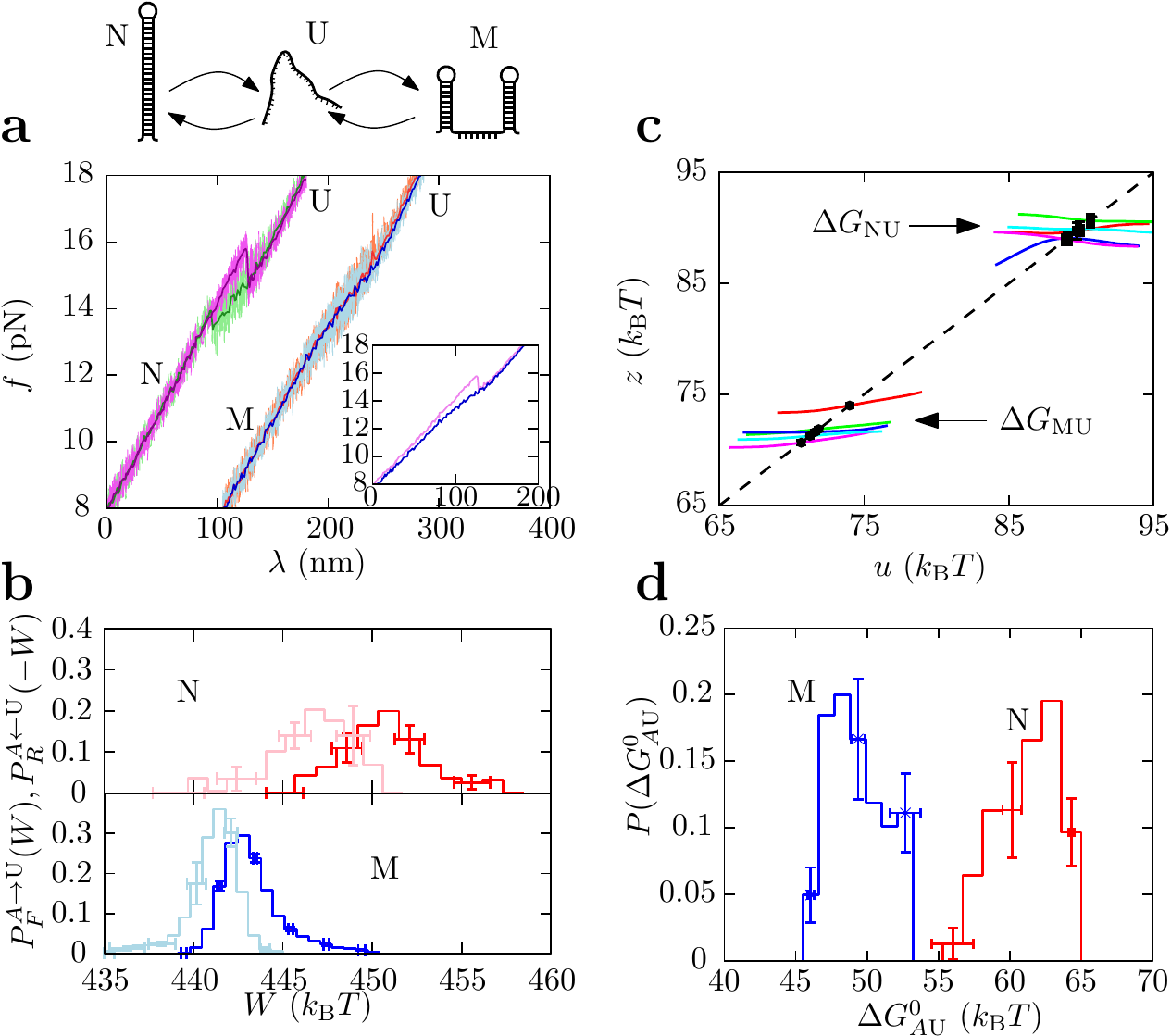}

\textbf{Hairpin M1, with one misfolded state.} 
\textbf{a)} Schematic illustration of the mechanical unfolding/folding pathways (top) and patterns identified in the FDC. For the sake of clarity curves are shifted.
Left: unfolding (purple) and folding (green) FDC for the N state show a force rip ($\simeq$ 0.5 pN) around 15 pN. 
Right: unfolding (red) and folding (blue) curves are the FDC for the M state.
In the analysis we used data from 5 molecules, pulled at 40, 90 and 125 nm/s. The number of unfolding/folding cycles performed varies between 80 and 270 for each dataset. 
\textit{Inset:} Unfolding curves for N and M without shift (data filtered). At low forces M has larger extension than N. 
\textbf{b)} Partial work histograms for work values measured at 125 nm/s between $\lambda_0$=0, where $A$=N (red, top panel) or M (blue, bottom panel) and $\lambda_1$=130 nm, where $B$=U. Dark colors refer to unfolding and light colors to folding work distributions.
\textbf{c)} The acceptance ratio method is applied to work values measured at different pulling speeds (40, 90 and 125 nm/s) to extract $\Delta G_{\rm NU}$ and $\Delta G_{\rm MU}$. Different colors refer to results obtained for different molecules pulled at different speeds.
\textbf{d)} Histograms of the free energy of formation of N (red) and M (blue) obtained for different molecules. Error bars in panels b, d indicate the standard statistical deviation computed over 270 cycles for a given molecule. These were obtained using the bootstrap method.

\subsubsection*{Fig. 4}

\includegraphics[scale=1]{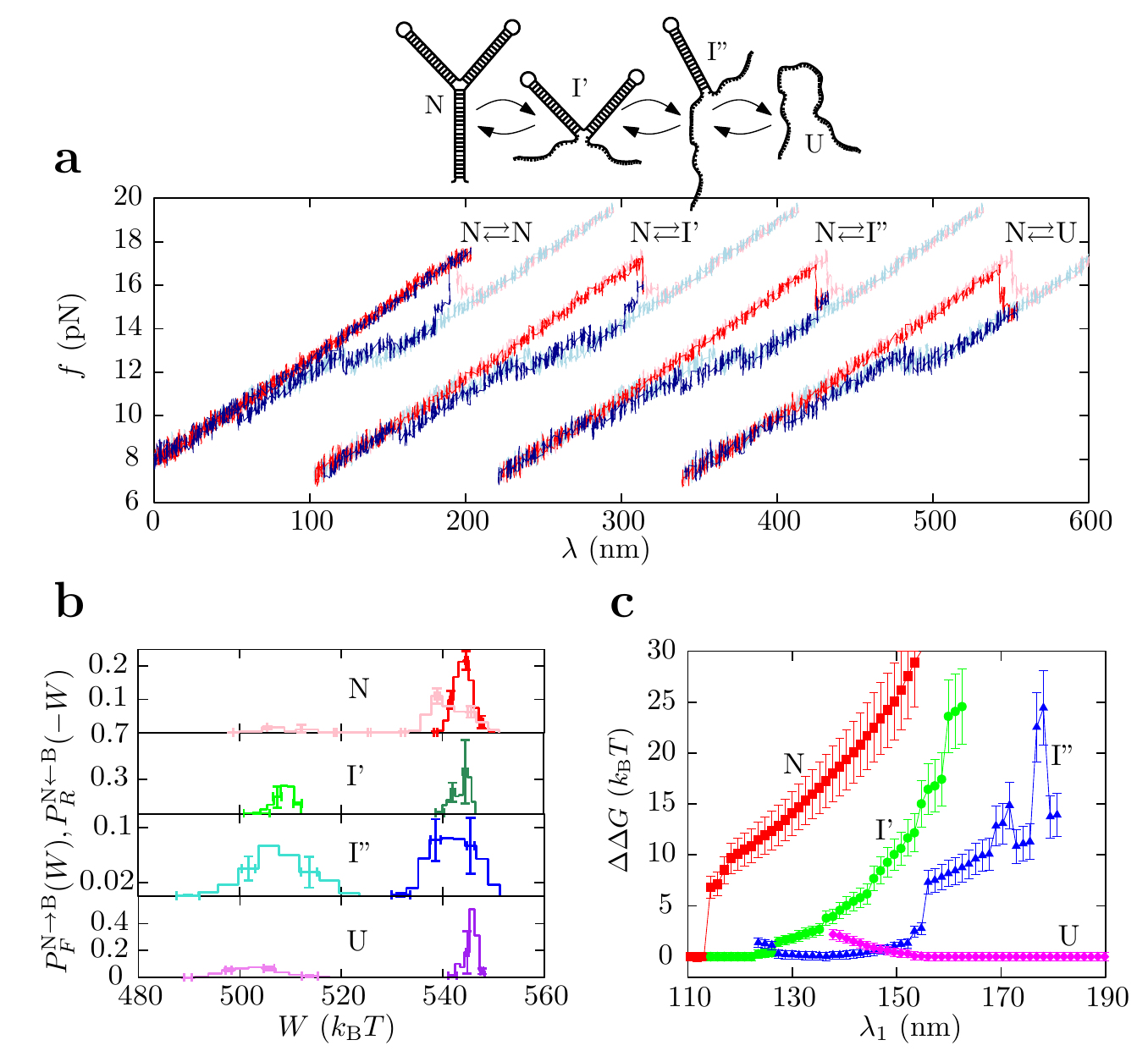}

\textbf{Hairpin I2, with two intermediate states.}
\textbf{a)} Schematic illustration of the mechanical
unfolding/folding pathway (top) and collection of unfolding (red) and folding (blue) FDCs recorded at
60 nm/s: the four branches of force correspond to the four states N, I', I'' and U. Curves plotted
using light colors are measured by setting extreme values of $\lambda$ such that the molecule equilibrates
at N and U at the initial and final pulling conditions respectively. In curves plotted using dark colors
$\lambda_1$ was chosen so that at the end of the F process and at the beginning of the R process the
molecule can be found in any state: N (two patterns are observed in the reverse trajectories with different degrees of dissipation), I', I'' or U. Curves corresponding to different transitions are shifted
for the sake of clarity. 7 molecules were pulled at 60 nm/s; between 100 and 400 cycles were
collected for each molecule. 
\textbf{b)} Partial work histograms for work values measured between $\lambda_0$=0,
where $A$=N, and $\lambda_1$=183 nm, where $B$=N (red, first panel), $B$=I' (green, second
panel), $B$=I'' (blue, third panel) or $B$=U (purple, bottom panel). Dark colors refer to forward work distributions and light colors to reversed work distributions. 
\textbf{c)} Free energy  branches of states N (red squares), I' (green circles), I'' (blue triangles) and U (purple diamonds) measured relative to the full free energy of the system, $\Delta G=-k_{\rm B}T\log \sum_{B={\rm N,I',I'',U}}e^{-\Delta G_{{\rm N}B}/k_{\rm B}T}$. Error bars in panels b, d indicate the standard statistical deviation computed over 400 cycles for a given molecule. These were obtained using the bootstrap method.

\subsubsection*{Fig. 5}

\includegraphics[scale=1]{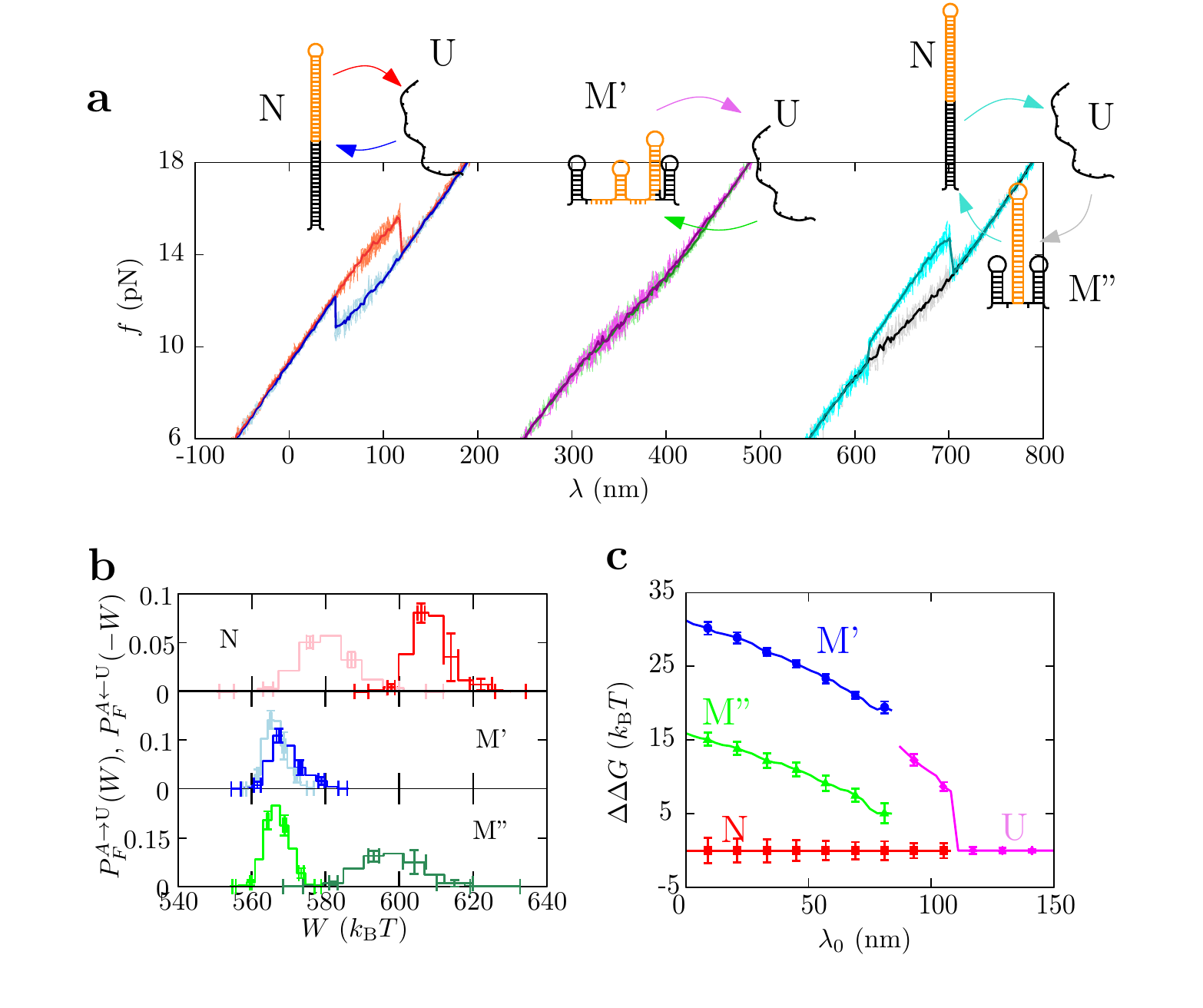}

\textbf{Hairpin M2, with two misfolded states.}
\textbf{a)} Patterns identified in the FDC with corresponding unfolding/folding pathways. N and M'' share a piece of hairpin in their folded conformation (orange).
Left: unfolding (red) and folding (blue) FDCs for state N show a force rip ($\simeq$2 pN) around 15 pN.
Middle: unfolding (purple) and folding (green) FDCs for state M' show no hysteresis.
Right: folding (gray) FDCs for state M'' are identical to the ones measured for M', whereas unfolding (cyan) FDCs show a rescue to the N state.
Six molecules were pulled, obtaining a minimum of 40 cycles and a maximum of 100 at 60 nm/s.
\textbf{b)} Partial work histograms for work values measured between $\lambda_0$=0, where $A$=N (red, top panel), M' (blue, middle panel) or M'' (green, bottom panel), and $\lambda_1$=230 nm, where $B$=U. Dark colors refer to unfolding and light colors to folding work distributions.
\textbf{c)} Free energy  branches of states N (red squares), M' (blue circles), M'' (green triangles) and U (purple diamonds) measured relative to the full free energy of the system, $\Delta G=-k_{\rm B}T\log \sum_{A={\rm N,M',M'',U}}e^{-\Delta G_{A{\rm U}}/k_{\rm B}T}$. Error bars in panels b, c indicate the standard statistical computed over 100 cycles for a given molecule. These were obtained using the bootstrap method.

\clearpage

\section*{Tables}

\begin{table}[ht]
\centering
\begin{tabular}{ccr@{$\pm$}lr@{$\pm$}lcc}
\hline
 & & \multicolumn{2}{c}{EFR} & \multicolumn{2}{c}{Hopping} & UO & Unzipping \\
\hline
I1 & $\Delta G_{\rm NI}^0$ & 30 & 3 & 31 & 2 & 30.5 & 27.1 \\
   & $\Delta G_{\rm NU}^0$ & 55 & 3 & 61 & 2 & 60.1 & 56.1 \\
\hline
M1 & $\Delta G_{\rm MU}^0$ & 47 & 2 & 46 & 3 &  49.6 & 46.9 \\
   & $\Delta G_{\rm NU}^0$ & 62 & 3 & 58 & 3 &  60.2 & 57.2 \\
\hline
I2 & $\Delta G_{\rm NI'}^0$  &  40& 6 & \multicolumn{2}{l}{\hspace{1.2em}-} &  41.9 & 39.4  \\
   & $\Delta G_{\rm NI''}^0$ &  80& 7 & \multicolumn{2}{l}{\hspace{1.2em}-} &  83.8 & 78.8  \\ 
   & $\Delta G_{\rm NU}^0$   & 125& 7 & \multicolumn{2}{l}{\hspace{1.2em}-} & 138.0 & 129.1 \\
\hline
M2 & $\Delta G_{\rm NU}^0$   & 94 & 2 & \multicolumn{2}{l}{\hspace{1.2em}-} &  92.9 & 87.4\\
   & $\Delta G_{\rm M'U}^0$  & 60 & 3 & \multicolumn{2}{l}{\hspace{1.2em}-} &  62.0 & 57.4\\
   & $\Delta G_{\rm M''U}^0$ & 70 & 3 & \multicolumn{2}{l}{\hspace{1.2em}-} &  72.3 & 67.9\\
\hline
\end{tabular}
\caption{Free energies of formation evaluated using the EFR and equilibrium-based methods (Supplementary Section S3) for the DNA sequences studied in this paper compared to predictions based on the nearest-neighbor (NN) model using the UO set of parameters \cite{Santalucia,Zuker2003} and data obtained from unzipping experiments \cite{HugBizForSmiBusRit10}. Error bars in the first and second columns contain the standard deviations over different molecules (statistics given in figure captions) and systematic errors in the calibration of the instrument (5\% in force and distance).
Among the two, the greatest contribution turns out to be
the force and distance calibration errors. These errors stem from appropriate conversion
factors between measured voltages in the detectors of forces and distances and are
multiplicative (Section S2 in \cite{HugBizForSmiBusRit10}). This means that our
free energy numbers have an absolute value as currently indicated here, and the
error in the ratio between experimental free energies is smaller.
Such error bars are compatible with the discrepancy observed between the different predictions provided by the NN model. An extension of this table is given in the Supplementary Section S8.
}
\end{table}

\clearpage

\clearpage

\def\thesection{S\arabic{section}} 
\def\theequation{S\arabic{equation}}
\def\thetable{S\arabic{table}}
\def\thefigure{S\arabic{figure}}

\section*{Supplementary Information}

\section{Evaluation of Free energy landscapes}

Each molecular structure (native or
misfolded) has a fixed number $N$ of basepairs in the folded state. 
Along the unfolding pathway, the molecule can explore multiple
conformations; their number grows exponentially with the number of
bases, and so grows the number of potentially stable intermediate
states. To model free energy landscapes (FEL) only sequential configurations are
taken into consideration \cite{Cocco2003}. Each configuration is 
labeled by the number of open basepairs $n$: $n$=0 corresponds to
the folded state (all basepairs are formed) and $n=N$ to the
fully unzipped unfolded state.

For a given value $n$ and a force $f$ the free energy is given by \cite{Forns2010}:

\begin{subequations}
\label{eq: FEL}
\begin{equation}
\label{eq: FEL-complet}
\Delta G_n(f)=\Delta G_n^0+\Delta G_n^{\rm ssDNA}(f)+\Delta G_n^d(f)
\end{equation}
\begin{align}
\label{eq: FEL-ssDNA}
\Delta G_n^{\rm ssDNA}(f)&=\int_0^{x_n(f)} F_{\rm ssDNA}^{l_n}(x')dx'-fx_n(f)\nonumber\\
&=-\int_0^f x_n(f')df'
\end{align}
\begin{align}
\label{eq: FEL-d}
\Delta G_n^d(f)&=\int_0^{x_d(f)} F_d(x')dx'-fx_d(f)\nonumber\\
&=-\int_0^f x_d(f')df'
\end{align}
\end{subequations}

where $\Delta G_n^0$ is the free energy of formation of the $n^{\rm th}$ configuration at zero force that can be estimated using the nearest-neighbor (NN) model and the unified-oligonucleotide (UO) set of parameters derived from bulk experiments \cite{Santalucia,Zuker2003} or data derived from single-molecule unzipping experiments \cite{JM-PNAS}.
$\Delta G_n^{\rm ssDNA}(f)$ is the elastic free energy at force $f$ of the released ssDNA for the configuration $n$, $x_n(f)$ being its equilibrium end-to-end distance projected along the force axis and $F_{\rm ssDNA}^{l_n}(x)$ the equation of state of a ssDNA polymer of contour length $l_n$. 
We use the inextensible worm-like chain (WLC) model with a persistence length $P=1.3\pm0.2$ nm and an inter-phosphate distance equal to 0.59 nm/base
\cite{JM-PNAS,EffectForce,DessingesCroquettePRL2002,WoodsideBlockPNAS2006,MossaJSTAT2010}.
$\Delta G_n^d(f)$ is the energetic contribution due to the orientation of the hairpin double helix (modeled as a single dipole of length $d$) along the force axis. Here we use an inextensible freely-jointed chain (FJC) model, with Kuhn and contour lengths equal to the diameter of the double helix,  $d\simeq$2.0 nm
\cite{DessingesCroquettePRL2002,WoodsideBlockPNAS2006}. 

If the structure under consideration is made of more than one hairpin (state M for M1;  N, I' and I'' for I2; or M' and M'' for M2) multiple sequential configurations can take place for an intermediate value of $n$. In this case, an exponential Boltzmann sum of the different contributions gives the free energy $\Delta G_n(f)$. 

\begin{figure}[ht]
 \centering
\includegraphics[scale=1]{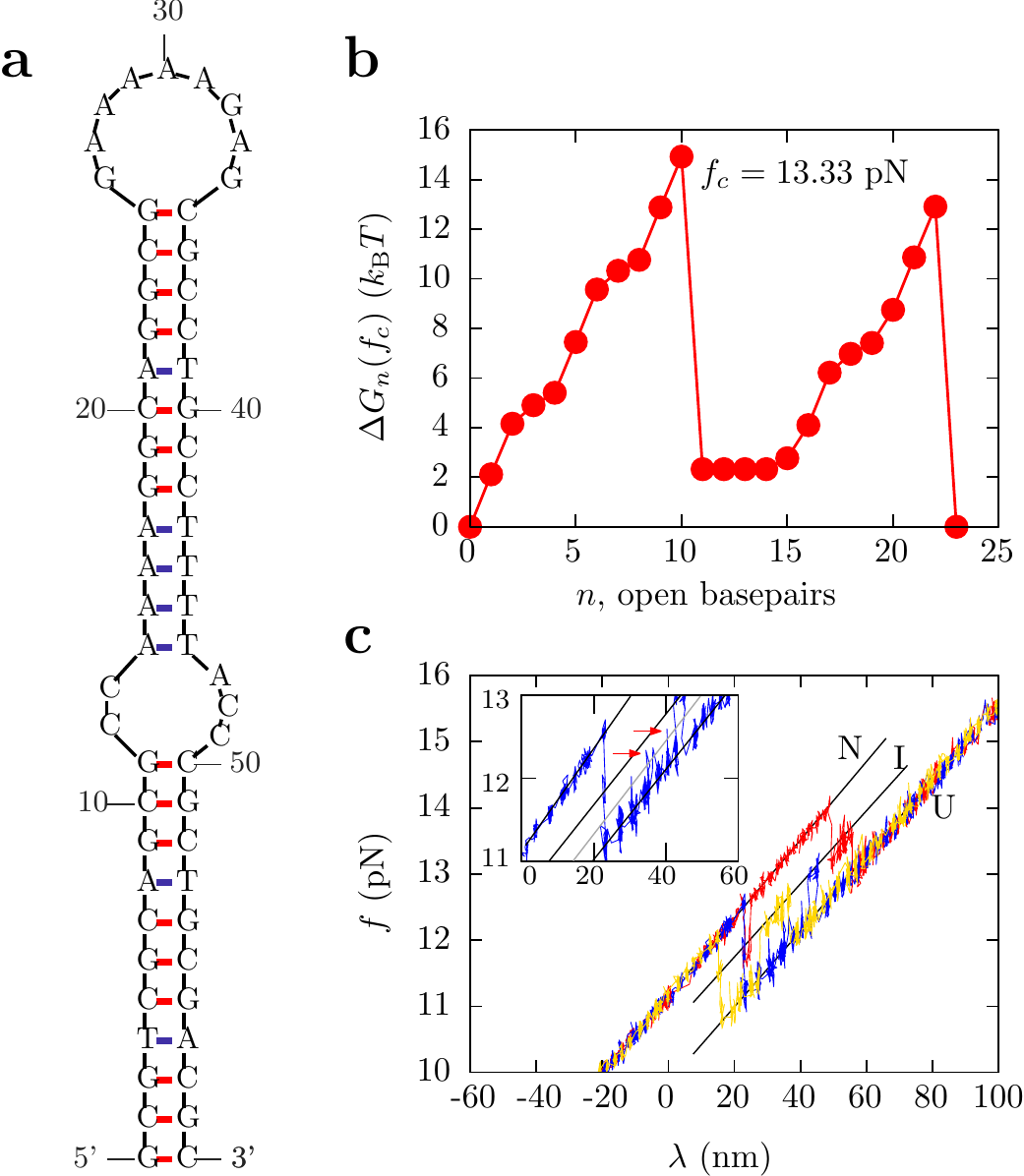}
\caption{\textbf{Molecule I1. Sequence, free energy landscape and FDC.}
\textbf{a.} Sequence and secondary structure. The native state of I1 is characterized by the presence of an internal loop in the middle of the stem that favors the existence of an intermediate (I) along the unfolding pathway of the hairpin.
\textbf{b.} Free energy landscape evaluated at the coexistence force (where the folded state, $n=0$, and the unfolded state, $n=N$, have equal free energies).  I contains four different configurations,  $n=11-14$, with the same free energy $\Delta G_n(f_c)$.
\textbf{c.} Example of unfolding FDC (red) and refolding FDC (blue,
yellow). The branches of force are indicated with black straight
state-lines. Each branch is assigned to one state of the molecule: N,
I or U. Inset: Detail of an unfolding curve where some points (indicated with red arrows) can be
assigned either to I or U depending on the size of the
running average used for the classification of the data points.
}\label{fig: S3}
\end{figure}

\begin{figure}[ht]
\centering
\includegraphics[scale=1]{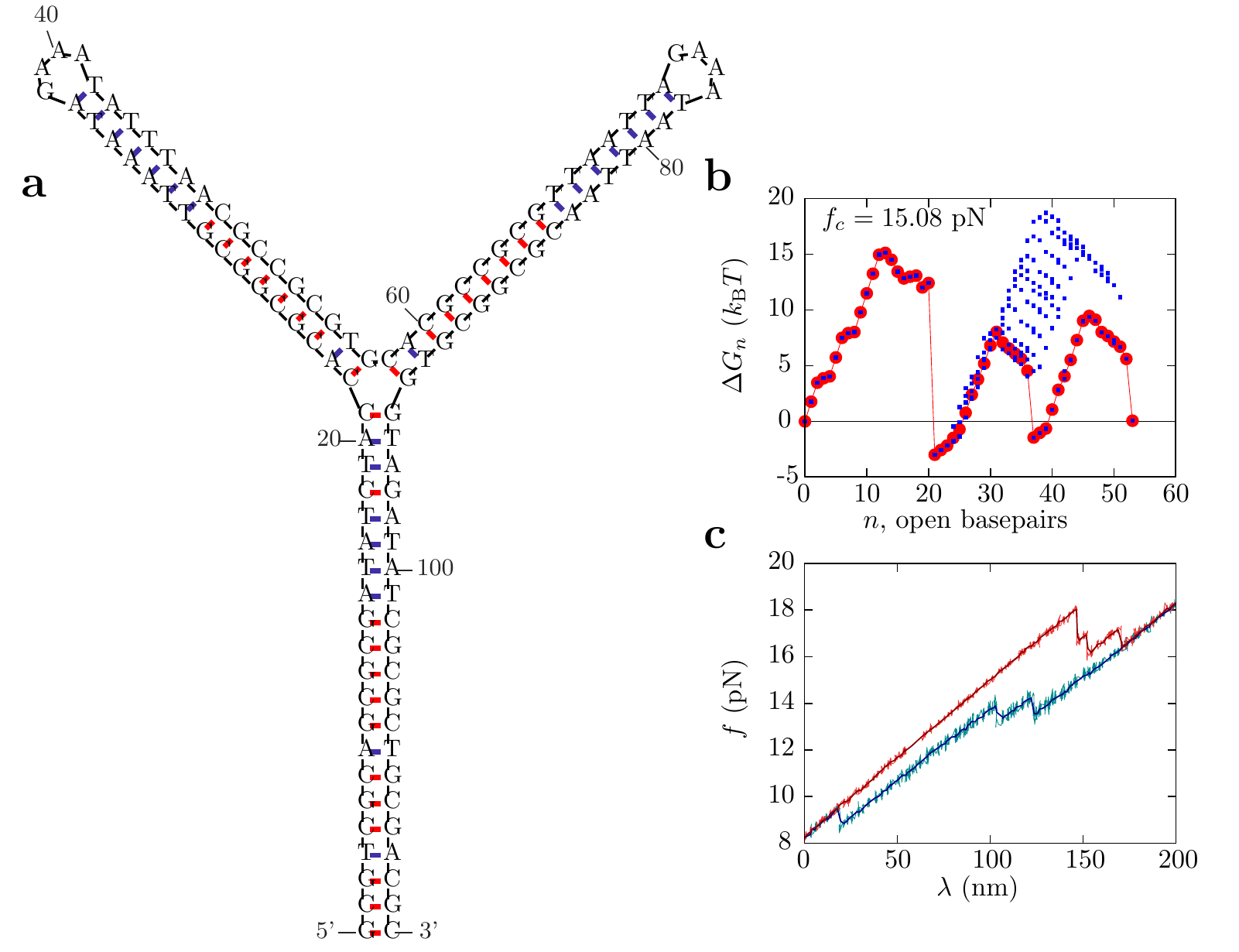}
\caption{\textbf{Molecule I2. Sequence, free energy landscape and FDC.} 
\textbf{a.} Sequence and secondary structure. The native state of I2 is characterized by the presence of a three way junction that favors the existence of two intermediates (I' and I'') along the unfolding pathway of the hairpin.
\textbf{b.} Free energy landscape calculated at the coexistence force (15.08 pN) predicted by the NN model using the UO parameters. 
Two intermediates, hereafter referred as I' and I'', appear along the unfolding pathway surrounded by high kinetic barriers (10 $k_{\rm B}T$). I' is found at $n=21$ and corresponds to a structure where all the stem is open and the two hairpins of the bifurcation are closed. I'' is found at $n=37$ and corresponds to two identical structures, where all the stem and one hairpin in the bifurcation are open. As this two configurations have identical molecular extension, we will consider I'' as a single state. 
Blue squares are the free energies of all the possible sequential configurations that have a given number of $n$ of unzipped base pairs distributed between the two hairpins. Red circles are the Boltzmann average (i.e. the mean free energy potential) taken by summing over all configurations constrained by a given number of unzipped base pairs n. 
\textbf{c.} Example of unfolding (red) and refolding (blue) FDC along one trajectory. The four branches of force are assigned to the four states N, I', I'' and U.  }\label{fig: I2}
\end{figure}

\begin{figure}[ht]
 \centering
\includegraphics[scale=1]{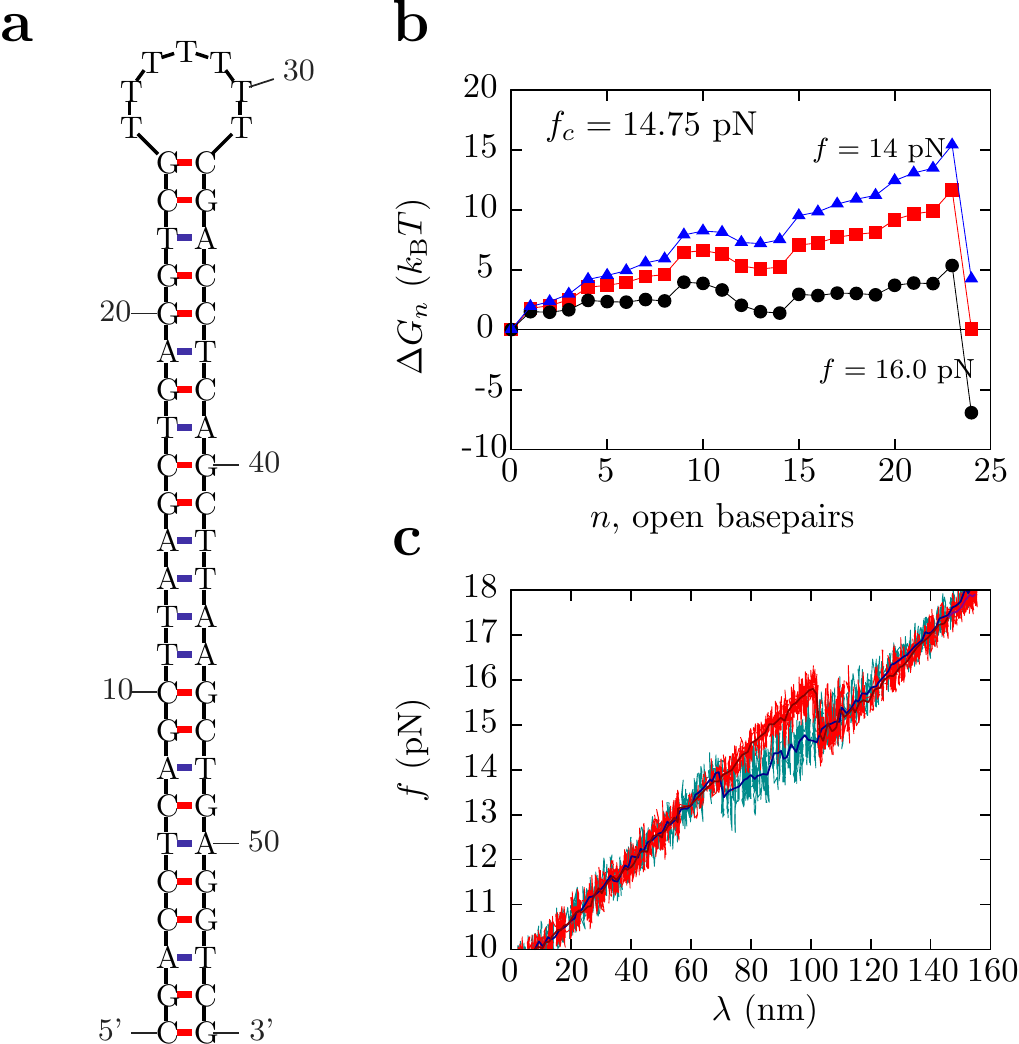}
\caption{\textbf{Molecule M1, N state. Sequence, free energy landscape and FDC.}
\textbf{a.} Sequence and secondary structure of N, which consists of a single-stem hairpin.
\textbf{b.} Free energy landscape evaluated at the coexistence force (14.75 pN, red squares), at 16 pN and at 14 pN. At the coexistence force an intermediate located at $n=13$ with a very low kinetic barrier to the native
state ($n=0$) can be seen. This intermediate may play a role in both unfolding and folding FDC (panel c), where force fluctuations increase in the vicinity of unfolding and folding rupture forces. For the sake of simplicity this intermediate is not characterized in our study. 
\textbf{c.} Example of unfolding (red) and refolding (blue) FDC.}\label{fig: M1N}
\end{figure}

\begin{figure}[ht]
\centering
\includegraphics[scale=1]{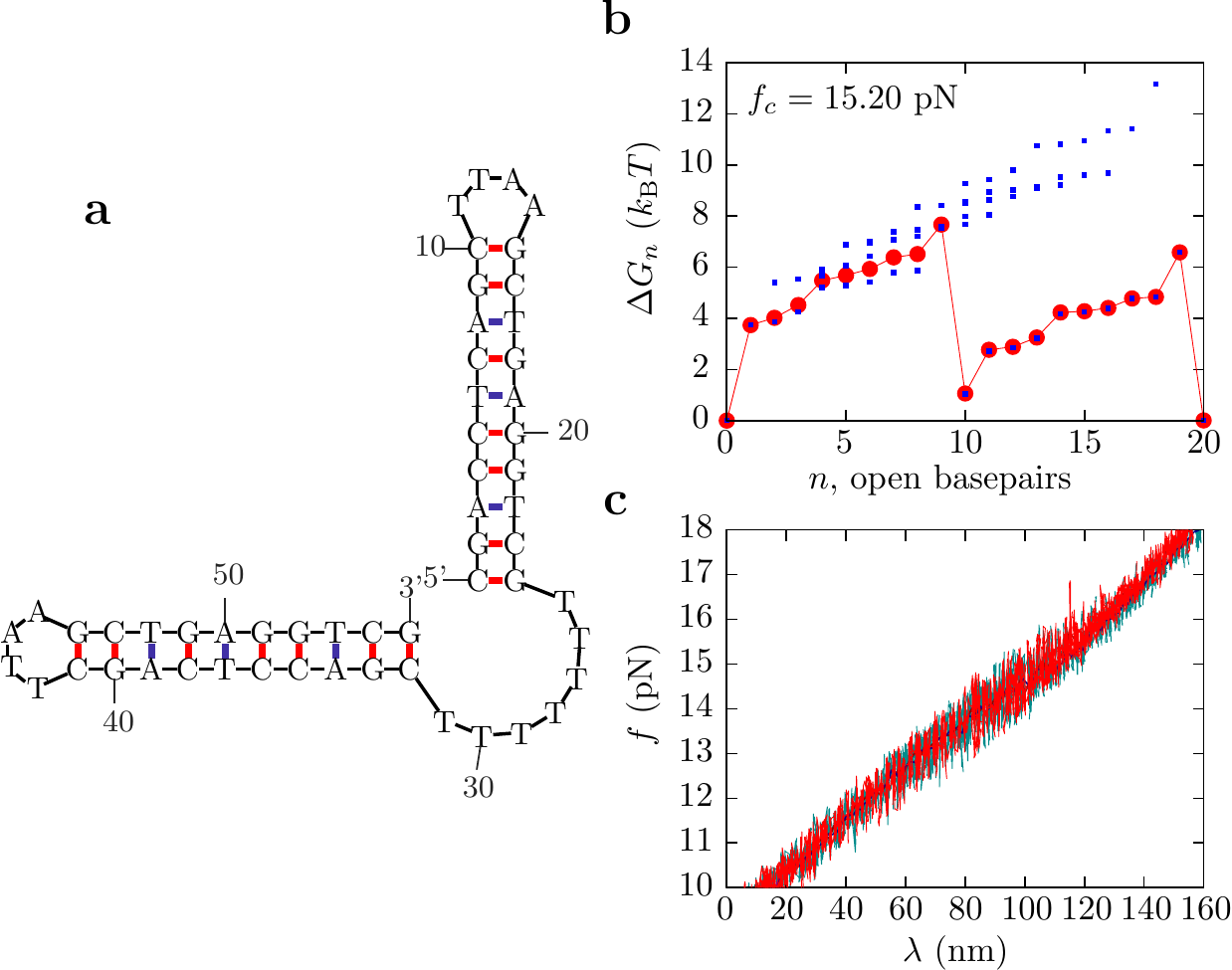}
\caption{\textbf{Molecule M1, M state.  Sequence, free energy landscape and FDC.}  
\textbf{a.} Sequence and secondary structure of M, which consists of two hairpins serially connected by seven thymines.
\textbf{b.} Free energy landscape evaluated at the coexistence force. Blue squares are the free energies
of all the possible sequential configurations of the structure. Red circles are the exponential Boltzmann average over configurations constrained by a given number of unzipped basepairs $n$.
An intermediate state surrounded by high kinetic barriers ($\sim$8 $k_{\rm B}T$) appears along the unfolding pathway at $n=10$. Configurations that dominate the exponential sum in the Boltzmann average for this value of $n$ are (10,0) and (0,10), where one of the hairpins is unzipped and the other is folded. For the sake of simplicity this intermediate is not characterized. 
\textbf{c.} Example of unfolding (red) and refolding
(blue) FDC.}\label{fig: M1M}
\end{figure}

\begin{figure}[ht]
\centering
\includegraphics[scale=1]{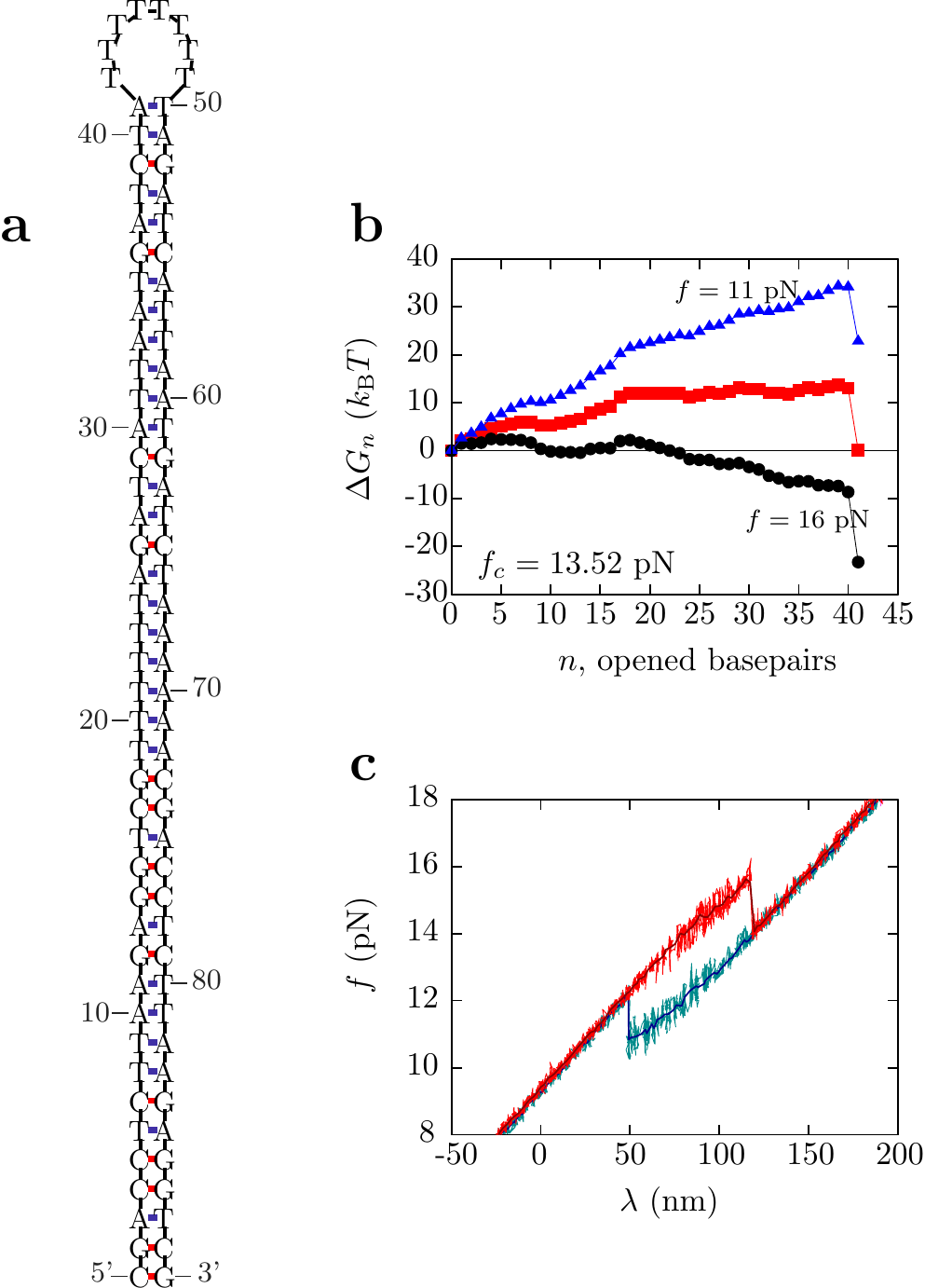}
\caption{\textbf{Molecule M2, N state.  Sequence, free energy landscape and FDC.}
\textbf{a.} Sequence and secondary structure of N.
\textbf{b.} Free energy landscape evaluated at the coexistence force (13.52 pN), 16 pN and 11 pN. 
\textbf{c.} Example of unfolding (red) and refolding (blue) FDC.}\label{fig: M2N}
\end{figure}

\begin{figure}[ht]
\centering
\includegraphics[scale=1]{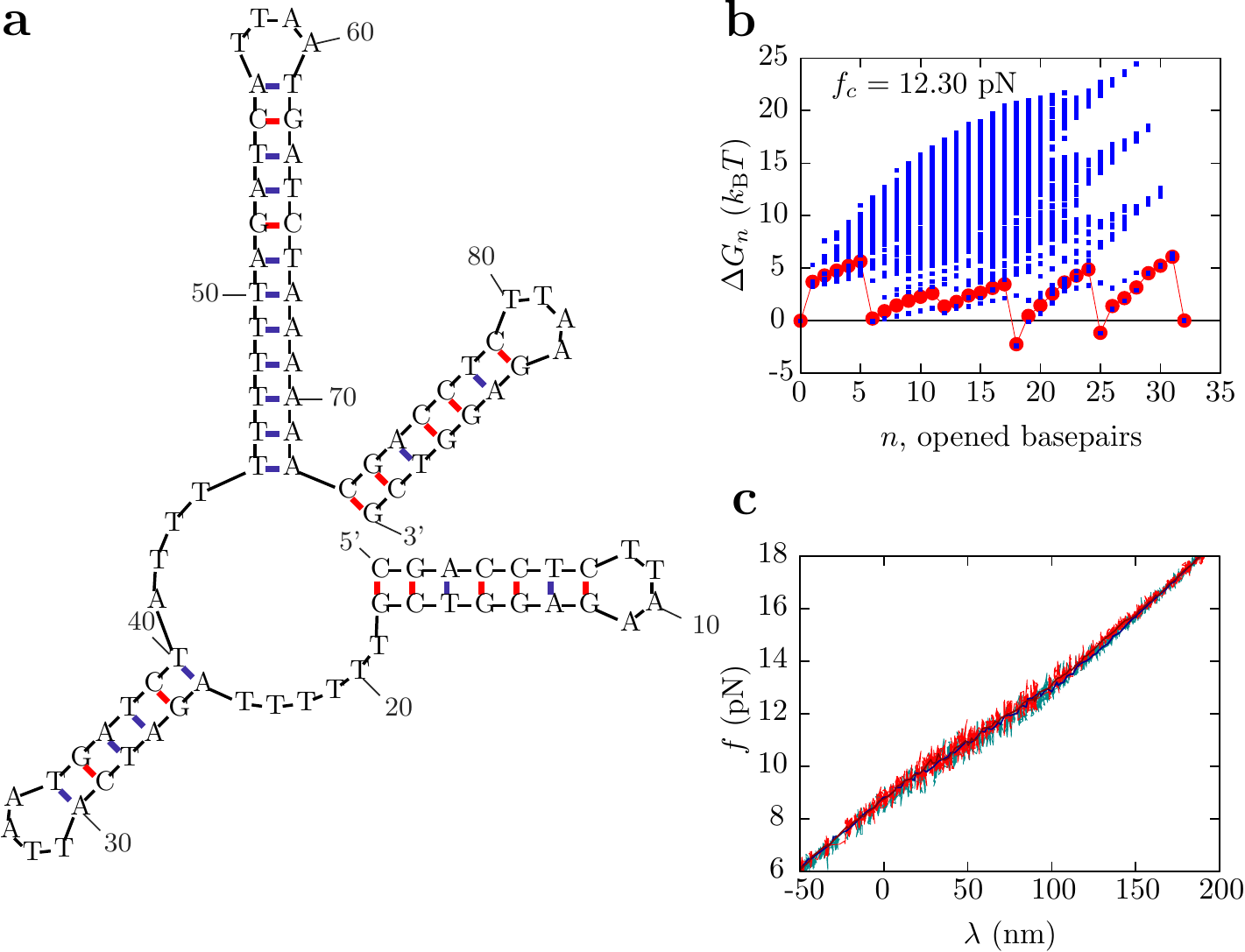}
\caption{\textbf{Molecule M2, M' state.  Sequence, free energy landscape and FDC.}
\textbf{a.} Sequence and secondary structure of M', which consists of four small hairpins that do not share any
structure with N .
\textbf{b.} Free energy landscape evaluated at the coexistence force. Blue squares are the free energies of all the possible sequential configurations of the structure. Red circles are the Boltzmann average taken over all configurations
constrained by a given number of unzipped base pairs $n$. Several intermediates separated by low kinetic barriers ($\sim$5 $k_{\rm B}T$) appear: one is located at $n=6$,
being the most stable configuration (0,6,0,0) where the first, third
and fourth hairpins are folded and the second hairpin is fully
unfolded. Other intermediates are found at $n=12$ (where (0,6,6,0) and
(0,0,12,0) dominate the Boltzmann average), $n=18$
(0,6,12,0) and $n=25$ ((7,6,12,0) and (0,6,12,7)).
\textbf{c.} Example of unfolding (red) and refolding (blue) FDC. The presence of
many transition states with low kinetic barriers implies an almost
reversible pattern for the FDC.}\label{fig: M2M2}
\end{figure}

\begin{figure}[ht]
\centering
\includegraphics[scale=1]{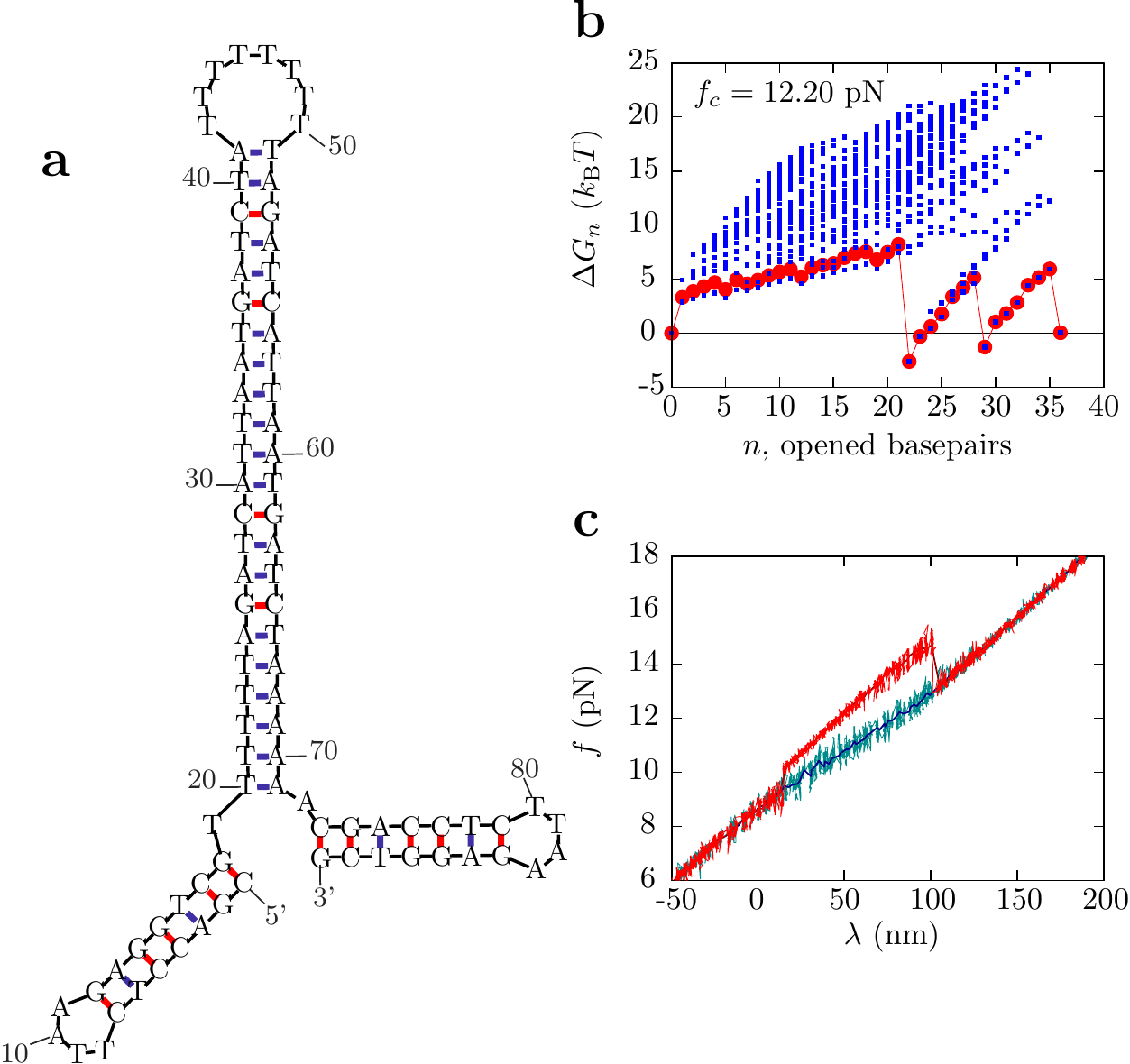}
\caption{\textbf{Molecule M2, M'' state.  Sequence, free energy landscape and FDC.}
\textbf{a.} Sequence and secondary structure of M'', which consists of three hairpins, one having 21 basepairs in common with N (Fig. \ref{fig: M2N}).
\textbf{b.} Free energy landscape evaluated at the coexistence
force. 
Blue squares are the free energies of all the possible sequential configurations of
the structure. Red circles are the Boltzmann average taken over all configurations
constrained by a given number of unzipped base pairs $n$. 
The FEL evaluated at the coexistence force reveals two intermediates at $n=22$ (configuration (0,22,0)) and $n=29$ (configurations (7,22,0) and (0,22,7)).
\textbf{c.} Example of unfolding (red) and refolding (blue) FDC.}\label{fig: M2M1}
\end{figure}

Following, the FEL of the different folded structures under consideration are shown. I1 and I2 fold only into one native structure and intermediates are revealed along the FEL (Figs. \ref{fig: S3} and \ref{fig: I2}). Molecule M1 can fold via two different pathways into two unrelated structures: the native (N, Fig. \ref{fig: M1N}) or the misfolded (M, Fig. \ref{fig: M1M}). Molecule M2 can fold into one native structure (N in Fig. \ref{fig: M2N}), and two different misfolded structures (M' in Fig. \ref{fig: M2M2} or M'' in Fig. \ref{fig: M2M1}).

% \subsection{Molecule I1}

% \subsection{Molecule I2}

% \subsection{Molecule M1}

% \subsection{Molecule M2}

\section{Free energy recovery from equilibrium experiments}

\subsection{Molecule I1}

\begin{figure}[ht]
\centering
\includegraphics[scale=1]{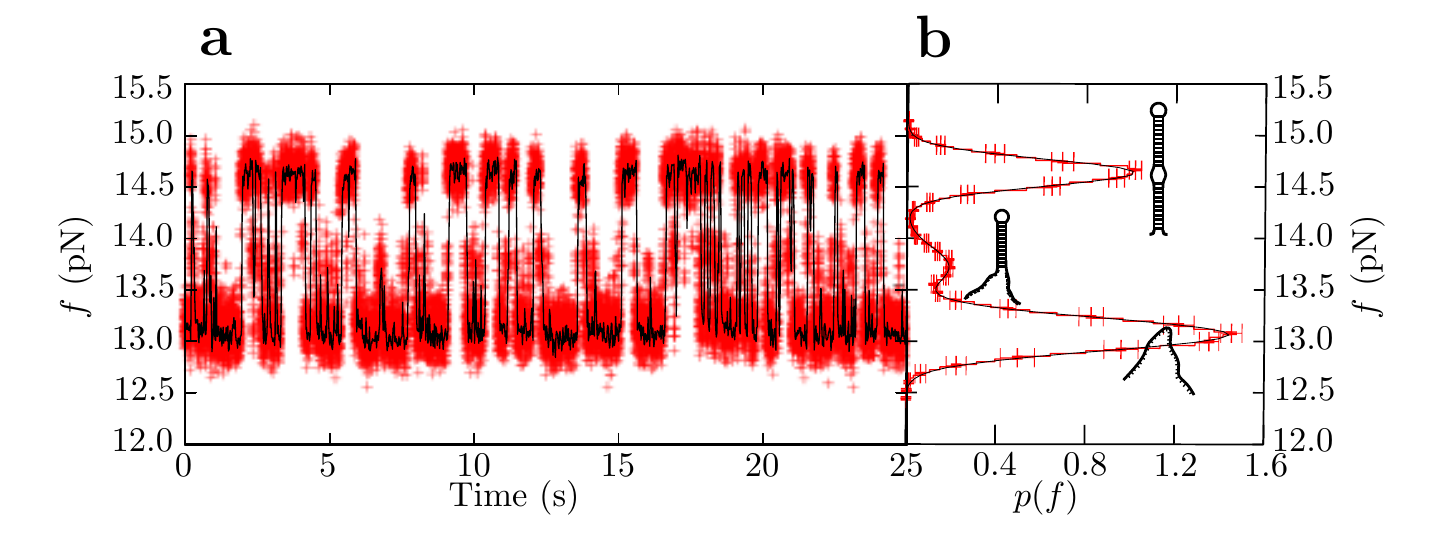}
\caption{\textbf{Hopping experiments with hairpin I1.} In these experiments the trap-pipette distance $\lambda$ is kept stationary. 
\textbf{a.} Sample trace during 25 seconds. In red we show data at full resolution and in black we show data filtered. The molecule executes transitions between N (at $\sim$14.5 pN), I (at $\sim$13.8 pN) and U (at $\sim$13 pN). 
\textbf{b.} Histogram of the measured forces during the hopping experiment. The free energy of formation of each state can be  obtained by fitting the histogram to a sum of three Gaussians and using the detailed balance condition, by relating the logarithm of the relative weights of the Gaussians to the free energy differences between the states.
}\label{fig: S3-eq}
\end{figure}

In the equilibrium-based hopping experiments the trap-pipette distance $\lambda$ is kept stationary and the molecule executes transitions between the three different states (N, I, U). A typical trace is shown in Fig. \ref{fig: S3-eq}a. 

The histogram of the measured force along an equilibrium trace can be fitted to a sum of three Gaussians:
\begin{equation}
 p(f)=\frac{w_N}{\sqrt{2\pi\sigma^2_N}}e^{-\frac{1}{2}\frac{(f-\langle f_N\rangle)^2}{\sigma^2_N}}+\frac{w_I}{\sqrt{2\pi\sigma^2_I}}e^{-\frac{1}{2}\frac{(f-\langle f_I\rangle)^2}{\sigma^2_I}}+\frac{w_U}{\sqrt{2\pi\sigma^2_U}}e^{-\frac{1}{2}\frac{(f-\langle f_U\rangle)^2}{\sigma^2_U}}
\end{equation}
where $w_N$, $w_I$ and $w_U$ are the relative weights of each state N, I and U respectively ($w_N+w_I+w_U=1$); $\langle f\rangle_N$, $\langle f\rangle_I$ and $\langle f\rangle_U$ are their average forces; and $\sigma_N$, $\sigma_I$ and $\sigma_U$ are the standard deviations. 
The free energy of formation of I and U (relative to N) can be obtained using the detailed balance condition, by relating the logarithm of the relative weights of the Gaussians to the free energy differences between states. To get the free energy at zero force we need to subtract the elastic contribution of each state (derived in section S2) at the average force of the equilibrium trace $\langle f\rangle$:

\begin{equation}
 \Delta G_{\rm NI}^0=-k_BT\log\left(\frac{w_I}{w_N}\right)+\int_0^{\langle f\rangle}x_I(f)df
\end{equation}
\begin{equation}
 \Delta G_{\rm NU}^0=-k_BT\log\left(\frac{w_U}{w_N}\right)-\int_0^{\langle f\rangle}x_d(f)df+\int_0^{\langle f\rangle}x_U(f)df
\end{equation}

Results, summarized in Table 1 in the main paper, are in agreement with free energy predictions using the NN model with the UO set of parameters and data from unzipping \cite{Santalucia,Zuker2003,JM-PNAS}.

\subsection{Molecule M1}

\begin{figure}[ht]
\centering
\includegraphics[scale=1.05]{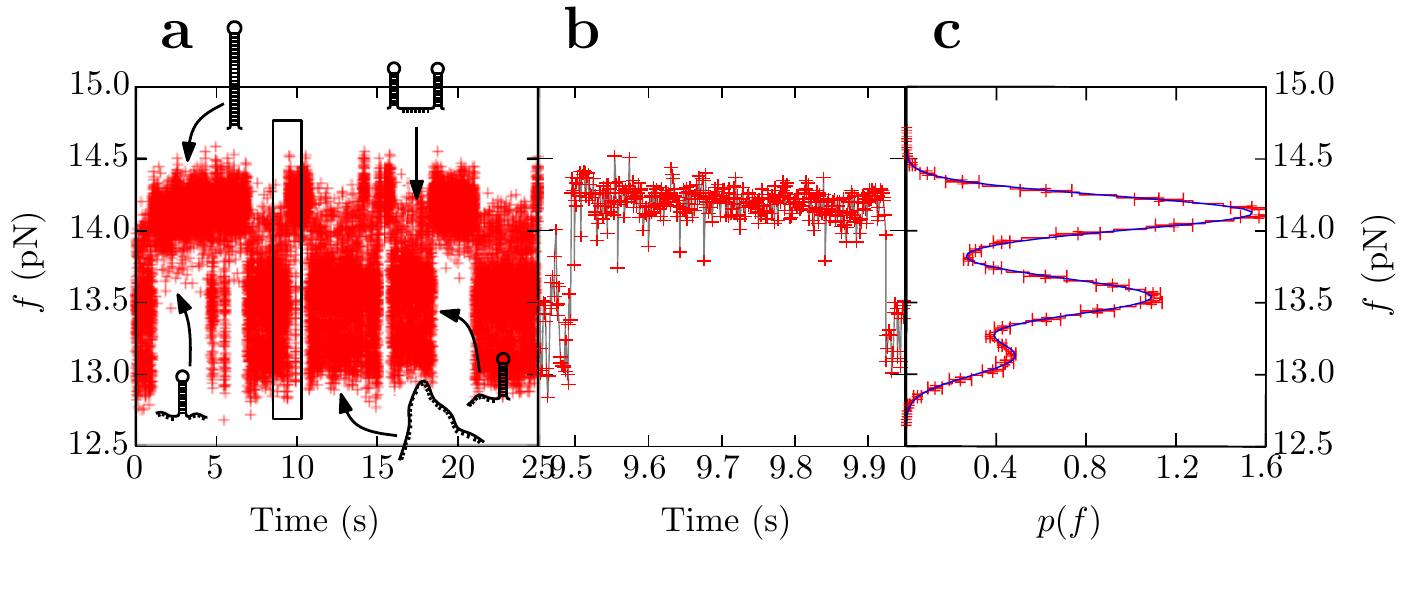}
\caption{\textbf{Hopping experiments with hairpin M1.} In these experiments the trap-pipette distance $\lambda$ is kept stationary. 
\textbf{a.} Sample trace at full resolution. The molecule executes transitions between N ($\sim$14.1 pN), M ($\sim$13.9 pN), two intermediate states I$_{\rm N}$ and I$_{\rm M}$ on-pathway to the native and to the misfolded state respectively ($\sim$13.5 pN in both cases since the molecular extension is identical) and U ($\sim$13.0 pN). 
\textbf{c.} Detail of the boxed region in panel a. A closer look of the traces at the interphase between both regions shows that there are not trajectories directly connecting N and M. 
\textbf{b.} Histogram of the measured forces during the hopping experiment. The free energy of formation of each state can be obtained using a fit to a sum of five Gaussians.}\label{fig: M1-eq}
\end{figure}

In the case of equilibrium experiments with M1, kinetics are very fast and some transitions are missed. 
Hopping traces in M1 reveal two clearly separated hopping regions (Fig. \ref{fig: M1-eq}): in one region states N and an intermediate I$_{\rm N}$ coexist; in the other region states M, U and another intermediate I$_{\rm M}$ coexist. A closer look of the traces at the interphase between both regions shows that there are no trajectories directly connecting N and M (Fig. \ref{fig: M1-eq}). 

The histogram of the measured force along an equilibrium trace can be fitted to a sum of five Gaussians:
\begin{align}
 p(f)&=\frac{w_N}{\sqrt{2\pi\sigma^2_N}}e^{-\frac{1}{2}\frac{(f-\langle f_N\rangle)^2}{\sigma^2_N}}+\frac{w_{I_N}}{\sqrt{2\pi\sigma^2_{I_N}}}e^{-\frac{1}{2}\frac{(f-\langle f_{I_N}\rangle)^2}{\sigma^2_{I_N}}}+\frac{w_U}{\sqrt{2\pi\sigma^2_U}}e^{-\frac{1}{2}\frac{(f-\langle f_U\rangle)^2}{\sigma^2_U}}\nonumber\\
&+\frac{w_{I_M}}{\sqrt{2\pi\sigma^2_{I_M}}}e^{-\frac{1}{2}\frac{(f-\langle f_{I_M}\rangle)^2}{\sigma^2_{I_M}}}+\frac{w_N}{\sqrt{2\pi\sigma^2_N}}e^{-\frac{1}{2}\frac{(f-\langle f_N\rangle)^2}{\sigma^2_N}}
\end{align}
where $w_i$, $i=$N, I$_N$, U, I$_M$, M are the relative weights of each state ($\sum_iw_i=1$); $\langle f\rangle_i$ are their average forces; and $\sigma_i$ are the standard deviations. 
The free energy of formation of N and M (relative to U) can be obtained using the detailed balance condition, by relating the logarithm of the relative weights of the Gaussians to the free energy differences between states. To get the free energy at zero force we need to subtract the elastic contribution of each state at the average force of the equilibrium trace $\langle f\rangle$:

\begin{equation}
 \Delta G_{\rm NU}^0=-k_BT\log\left(\frac{w_{ U}}{w_{\rm N}}\right)-\int_0^{\langle f\rangle}x_d(f)df+\int_0^{\langle f\rangle}x_U(f)df
\end{equation}
\begin{equation}
 \Delta G_{\rm MU}^0=-k_BT\log\left(\frac{w_{ U}}{w_{\rm M}}\right)-2\int_0^{\langle f\rangle}x_d(f)df-\int_0^{\langle f\rangle}x_M(f)df+\int_0^{\langle f\rangle}x_U(f)df
\end{equation}

Results, summarized in Table 1 in the main paper, are in agreement with free energy predictions using the NN model with the UO set of parameters and data from unzipping \cite{Santalucia,Zuker2003,JM-PNAS}. Intermediate states are not characterized.

\subsection{Molecule I2}

\begin{figure}[ht]
\centering
\includegraphics[scale=1]{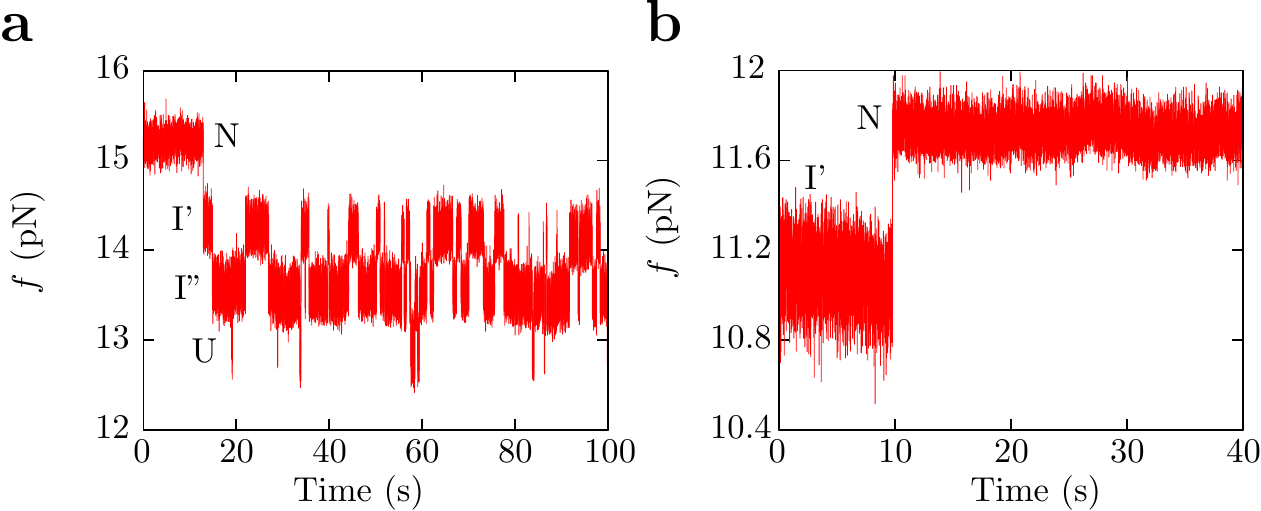}
\caption{\textbf{Pulling and hopping experiments with hairpin I2.}
\textbf{a.}  Example of a passive hopping trace measured at high forces: the molecule is initially set to N. Once the molecule partially unfolds (after $\sim$15 in the figure), the system subsequently hops between states I', I'' and U, but never folds back to N. 
\textbf{b.} Example of a hopping experiment at low forces: The molecule is initially set to I' (it was not possible to stabilize the molecule in neither U nor I''  states at such low forces) and once the molecule folds to N it never escapes out. Equilibrium hopping experiments were repeated at different forces and the native state was never observed to coexist with any other state. }\label{fig: I2-eq}
\end{figure}

Equilibrium-based hopping experiments reveal four levels of force, corresponding to the four different states N, I', I'' and U. Hopping between all four conformations was never observed in the accessible experimental time scales: at high forces hopping between I', I'' and U (but not N) occurs (Fig. \ref{fig: I2-eq}a); at low forces once the molecule reaches N it never leaves that conformation precluding equilibrium based free energy measurements (Fig.\ref{fig: I2-eq}b).

In the absence of coexistence between the four states the free energy of formation of I', I'' and U (relative to N) was not evaluated using equilibrium-based hopping experiments. 

\subsection{Molecule M2}

\begin{figure}[ht]
 \centering
\includegraphics[scale=0.7]{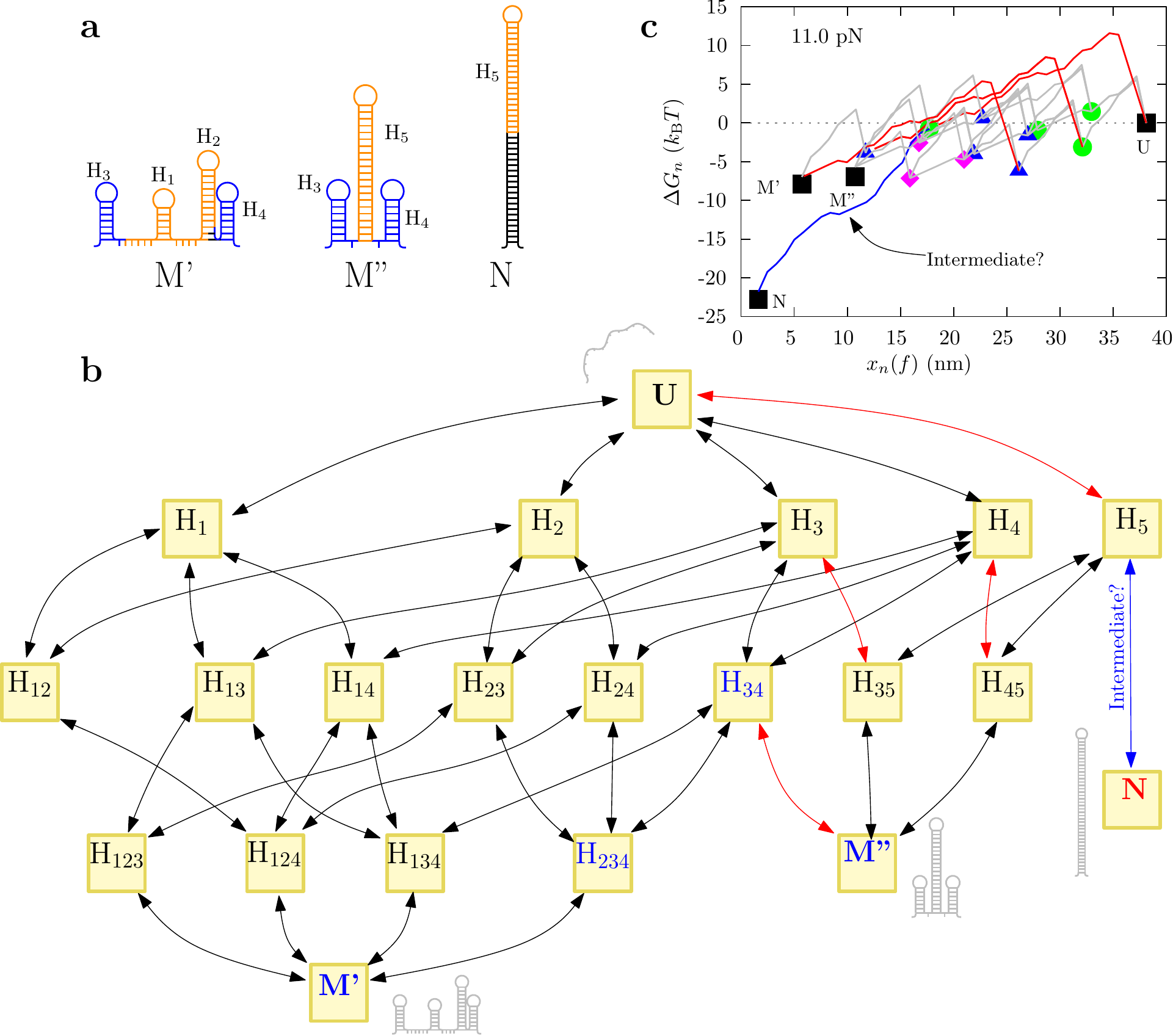}
\caption{\textbf{Folding pathways for N, M' and M''.} 
\textbf{a)} Folded structures for M2 according to Mfold \cite{Santalucia,Zuker2003}: M' is made by four small hairpins hereafter referred as H$_1$, H$_2$, H$_3$ and H$_4$; M'' is made by
hairpins H$_3$ and H$_4$ (also present in M'), and hairpin H$_5$; and N is made by a single and long hairpin
that contains H$_5$. 
\textbf{b)} The folding pathway to each state can be modeled by the sequential folding of H$_1$, H$_2$, H$_3$, H$_4$ or H$_5$. Starting from U a first
hairpin may form (H$_i$, $i$=1,2,3,4 or 5). Depending on the first hairpin, a second hairpin may form (H$_{ij}$=H$_i$+H$_j$). Folding proceeds to the next level of structures (H$_{ijk}$) down to the final states M', M'', N.
Black (red) arrows denote pathways that need to overcome a kinetic barrier of $\sim$5 ($\sim$10) $k_{\rm B}T$. The
blue arrow denotes the transition N$\rightleftarrows$H$_5$ , which might be mediated by an intermediate on-pathway
predicted in the free energy landscape. 
\textbf{c)} Free energy landscape along the different folding
pathways calculated at a constant force of 11 pN using the NN model and the UO set of
parameters. Folding pathways with barriers that are higher than 10 $k_{\rm B}T$ are plotted in red, and the
transition towards N is plotted in blue. Green circles denote first-level configurations of the type H$_i$,
blue triangles denote second-level configurations of the type H$_{ij}$ , purple diamonds denote third-level
configurations of the type H$_{ijk}$ and black squares indicate the initial state U and the final states N, M'
and M''.}\label{fig: M2-paths}
\end{figure}

The network of intermediate states present in the folding pathways of N, M' and M'' taking into account only sequential configurations is extremely complex (Fig. \ref{fig: M2-paths}). In fact, the molecule can change state from M' or M'' to N without going through U. 

\clearpage

\clearpage

\begin{figure}[ht]
\centering
\includegraphics[scale=1]{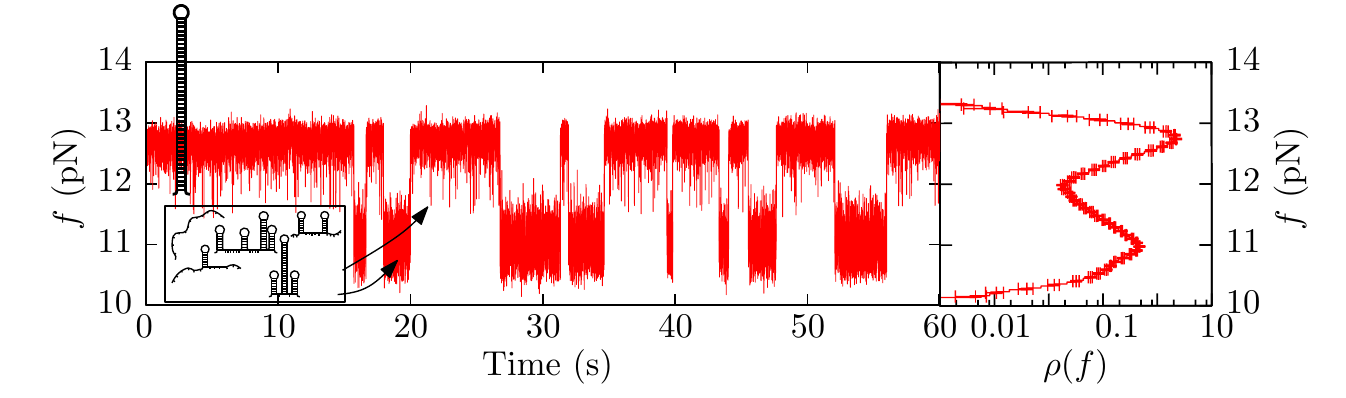}
\caption{\textbf{Hopping experiments for hairpin M2.}
At an intermediate range of forces (10-13pN) the hairpin shows hopping between N (mean force ~12.7 pN), an intermediate state (mean force ~11.7 pN) and a mixture several states (~mean force 11pN). The latter mixture is potentially composed of U, M', M'' and other intermediates (boxed region, see also Fig. \ref{fig: M2-paths}) that exhibit nearly the same molecular extension being very difficult to distinguish (rectangular box).}\label{fig: M2-equilibri}
\end{figure}

In the case of M2 kinetics in equilibrium experiments are not very fast (Fig. \ref{fig: M2-equilibri}) but it is very difficult to identify different conformations and unfolding/folding pathways along a hopping trace (due to the nearly identical molecular extensions of some states). In addition, as shown in Fig. \ref{fig: M2-equilibri}, the complexity of the network of states for M2 is such that there are pathways connecting N, M' and M'' that do not go through U. 

Therefore, it is not possible to measure the free energies of N, M' and M'' (relative to U) and to unravel unfolding pathways from equilibrium-based hopping experiments.

\section{Identification of folding/unfolding pathways along FDCs}

In order to prove that molecules fold into the structures summarized in Figs. \ref{fig: S3}-\ref{fig: M2M1} we measure the molecular extension released/absorbed, $\Delta x_m$, along FDC as:
\begin{equation}
 \Delta x_m=\dfrac{\Delta f}{k_{\rm eff}},
\end{equation}
where $\Delta f$ is the change in force along the transition and $k_{\rm eff}$ is the effective stiffness of the molecular construct (dsDNA handles plus optical trap), measured as the slope of the FDC before the transition occurs. In Tables \ref{tab: S3ext}-\ref{tab: M2ext} the experimentally measured $\Delta x_m^{\rm exp}$ for the different molecules are compared to predictions obtained using the worm-like-chain (WLC) model with a persistence length equal to 1.35 nm. Results match theoretical predictions.

\vspace{0.5cm}

\begin{table}[ht]
 \centering
\begin{tabular}{cccc}
\hline
 & $\Delta f$ (pN) & $\Delta x_m^{\rm exp}$ (nm) & $\Delta x_m^{\rm WLC}$ (nm) \\
\hline
N$\rightarrow$I & 0.76$\pm$0.04 & 12$\pm$1 & 12.0\\
I$\rightarrow$I & 0.57$\pm$0.03 & 11$\pm$1 & 11.2\\
\hline
\end{tabular}
\caption{\textbf{Change in molecular extension measured for different transitions in hairpin I1.}
Results are in agreement with predictions obtained using the WLC model. Statistical errors are
insignificant compared to systematic error, which we take equal to 5\%.} \label{tab: S3ext}
\end{table}

\begin{table}[ht]
 \centering
\begin{tabular}{cccc}
\hline
 & $\Delta f$ (pN) & $\Delta x_m^{\rm exp}$ (nm) & $\Delta x_m^{\rm WLC}$ (nm) \\
\hline
N$\rightarrow$U & 1.20$\pm$0.06 & 24$\pm$1 & 23.0\\
M$\rightarrow$U ($^*$) & 0.80$\pm$0.08 & 16$\pm$2 & 18.1\\
\hline
\end{tabular}
\caption{\textbf{Change in molecular extension measured for different transitions in hairpin M1.}
Results are in agreement with predictions obtained using the WLC model. Statistical errors are
insignificant compared to systematic error, which we take equal to 5\%. ($^*$) FDCs do not show a
sudden jump in force for the transition M$\rightarrow$U. $\langle\Delta f\rangle$ was then calculated from the relative shift
between each force branch M and U (see the inset of Fig. 3a at low forces in main text).} \label{tab: M1ext}
\end{table}

\begin{table}[ht]
 \centering
\begin{tabular}{cccc}
\hline
 & $\Delta f$ (pN) & $\Delta x_m^{\rm exp}$ (nm) & $\Delta x_m^{\rm WLC}$ (nm) \\
\hline
N$\rightarrow$I'   & 1.30$\pm$0.07 & 20$\pm$1 & 21.1\\
I'$\rightarrow$I'' & 0.82$\pm$0.05 & 13$\pm$1 & 14.7\\
I''$\rightarrow$U  & 0.80$\pm$0.05 & 14$\pm$1 & 14.7\\
\hline
\end{tabular}
\caption{\textbf{Change in molecular extension measured for different transitions in hairpin I2.}
Results are in agreement with predictions obtained using the WLC model. Statistical errors are
insignificant compared to systematic error, which we take equal to 5\%.
} \label{tab: I2ext}
\end{table}

% \clearpage

\begin{table}[ht]
 \centering
\begin{tabular}{cccc}
\hline
 & $\Delta f$ (pN) & $\Delta x_m^{\rm exp}$ (nm) & $\Delta x_m^{\rm WLC}$ (nm) \\
\hline
N$\rightarrow$U        & 1.50$\pm$0.08 & 35$\pm$2 & 38.5\\
M''$\rightarrow$N      & 0.80$\pm$0.05 & 18$\pm$1 & 15.9\\
M$\rightarrow$U ($^*$) & 1.3$\pm$0.2 & 28$\pm$4 & 30.8\\
\hline
\end{tabular}
\caption{\textbf{Change in molecular extension measured for different transitions in hairpin M2.}
Results are in agreement with predictions obtained using the WLC model. Statistical errors are
insignificant compared to systematic error, which we take equal to 5\%. ($^*$) FDCs do not show a
sudden jump in force for the transition M$\rightarrow$U. $\langle\Delta f\rangle$ was calculated from the relative shift
between each force branch M and U (as in M1), and
values of for M' and for M''.} \label{tab: M2ext}
\end{table}

\clearpage

\section{Validation of EFR}

From the extended fluctuation relation (EFR) \cite{Junier} we can write:
\begin{equation}
\log\left(\frac{P_F^{A \rightarrow B}(W)}{P_R^{A\leftarrow B}(-W)}\right)
=\frac{W}{k_{\rm B}T}-\frac{\Delta G_{AB}}{k_{\rm B}T}-\log\left(\frac{\phi_F^{A\rightarrow B}}{\phi_R^{A\leftarrow B}}\right)
\end{equation}

\begin{figure}[ht]
\centering
\includegraphics[scale=0.8]{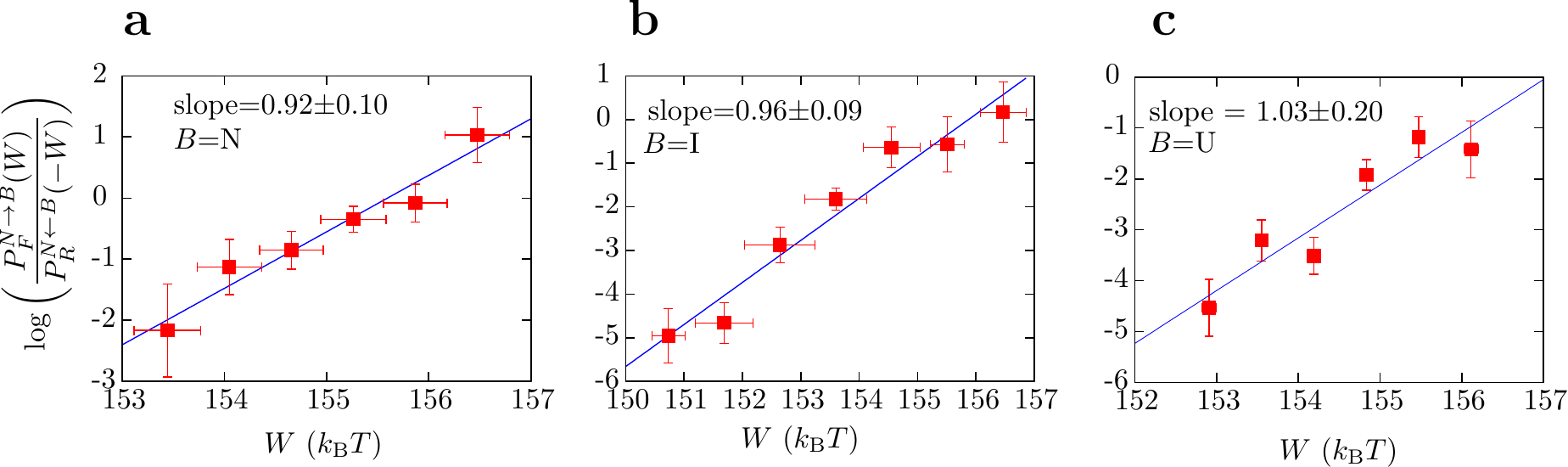}
\caption{\textbf{EFR test for I1.} Validation of the EFR using the partial work distributions shown in Fig. 2 for states N \textbf{(a)}, I \textbf{(b)} and U \textbf{(c)}. Error bars were obtained using the bootstrap method.
}\label{fig: SI-S3-lp}
\end{figure}

If we plot $\log\left(\frac{P_F^{A \rightarrow B}(W)}{P_R^{A\leftarrow
B}(-W)}\right)$ as a function of the work in $k_{\rm B}T$ units we
obtain a straight line with slope equal to 1 as a direct proof of the
validity of the EFR. Fig. \ref{fig: SI-S3-lp} show
experimental tests of the validity of the EFR for molecule I1 taking $A$=N and $B$=N, I or U.
The partial work distributions used are
shown in Fig. 2b in the main paper. Within experimental
errors the EFR holds for the three different kinetic states.

% \clearpage

\section{Potential sources of error: classification of states along FDCs}

\subsection{Molecules with intermediates on-pathway}

Measured FDC for molecules I1 and I2 show different force branches (Figs. \ref{fig: S3}c and \ref{fig: I2}c). Each branch corresponds to one state. 
By representing each branch with a characteristic straight
state-line (black lines in Fig. \ref{fig: S3}c) it is possible to
assign a state to each measured data point $(\lambda,f)$. This is done by
determining the nearest state-line to each measured
point. The drawback of this method stems from large force
fluctuations (Fig. \ref{fig: S3}c, inset): due to the finite acquisition rate, 1 kHz, sometimes there are
isolated points that can be assigned to the wrong state.
As a consequence, an error can be introduced in the fraction $\phi_F^{A\rightarrow B}$ and in the partial work distributions $P_F^{A \rightarrow B}(W)$ and $P_R^{A\leftarrow B}(-W)$, where $A$=N and $B$=N, I or U ($\phi_R^{A\leftarrow B}=1$ since at the chosen value of $\lambda_0$ all reversed trajectories end in state N).

\begin{figure}[ht]
 \centering
\includegraphics[scale=1]{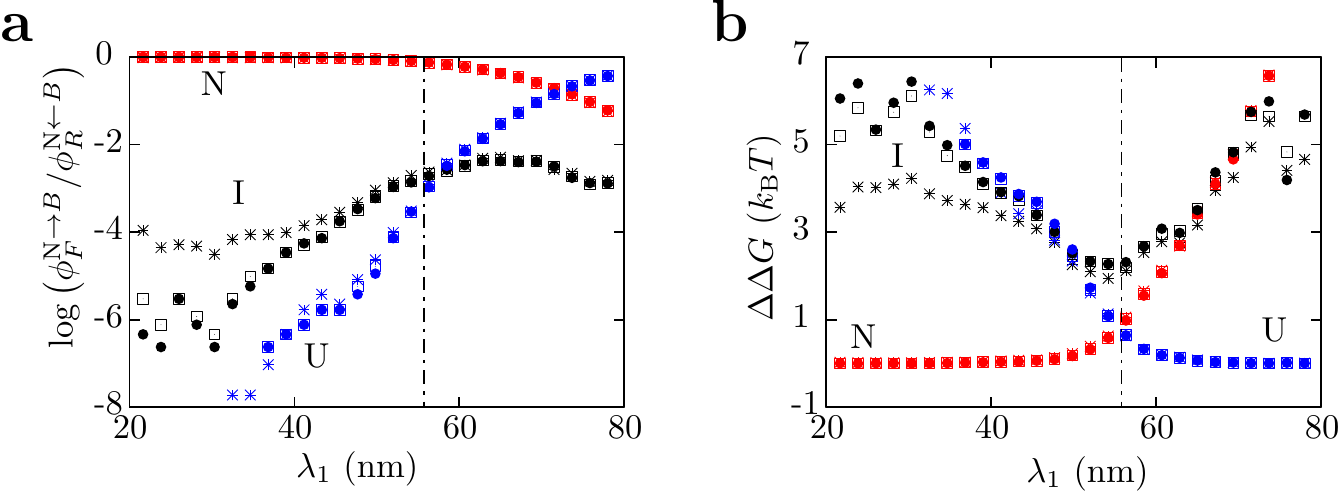}
\caption{\textbf{Molecule I1. Effect of running boxcar averages on state
classification.} 
\textbf{a.} The quantity $\log\phi_F^{{\rm N}\rightarrow
B}/\phi_R^{{\rm N}\leftarrow B}$ as a function of $\lambda_1$ is evaluated for different sizes of
the running average of experimental raw data: 1 point (asterisks), 5 points (open
squares) and 10 points (solid circles). $B=$ N for red, $B=$ I for black and $B=$ U for blue symbols.  
\textbf{b.} Difference between the free energy of each state
$\Delta G_{{\rm N}B}$ and the full free energy, $\Delta G=-k_{\rm B}T\log\sum_{B={\rm N, I, U}}e^{-\Delta G_{{\rm N}B}/k_{\rm B}T}$, evaluated at different values of $\lambda_1$ for different sizes of the running average.
Color code as in a.
}\label{fig: S3-p}
\end{figure}

In order to study this effect, different sizes of a running average (1, 5 or 10 points) of experimental data for molecule I1 are considered. Results for the prefactor $\log\left(\phi_F^{A\rightarrow B}/\phi_R^{A\leftarrow B}\right)$ evaluated taking $\lambda_0$=0 nm ($A$=N) and varying $\lambda_1$ between 20 and 80 nm can be seen in Fig. \ref{fig: S3-p}a for $B$=N (red), I (black) and U (blue). A convergence is observed for running averages between five and ten points. The error made in the evaluation of free energies can be large for intermediate states under short lifetime conditions. For instance, in the case of I1 an error of the order of 2 $k_{\rm B}T$ is made only for I at low values of $\lambda_1$.

\clearpage

\subsection{Molecules with misfolded states off-pathway}

From the whole FDC pattern we can identify if the hairpin has folded into states N or M. Therefore, no error is made when evaluating $\phi_R^{A\leftarrow B}$ ($\phi_F^{A\rightarrow B}$=1 always) for a high value of $\lambda_1$ (where molecule is unfolded) and a low value of $\lambda_0$ (where molecule is folded, either in N or M). 

In the case of M2 we assume that unfolding curves starting
from a misfolded state and showing rescue to N around 9.5 pN (Fig. \ref{fig: M2M1}c) start in M'' \cite{LiBusTin07}. 
Folding curves preceding unfolding curves that are rescued by N are assigned to fold into M''; and folding curves preceding unfolding curves without rescue are assigned to fold into M'.
A crucial question to justify this assumption is how far from equilibrium the system is: if we are under quasi-static conditions the molecule has more time to explore the free energy landscape and to overcome high kinetic barriers, preferentially folding into N. On the other hand, if we are far from equilibrium the molecule will mostly fold into M' because its folding pathway encounters lower kinetic barriers.  In fact, as shown in the folding network of M2 sketched in Fig. \ref{fig: M2-paths}, the molecule can change conformation from M'' to N via two intermediates (H$_{45}$ or H$_{35}$ and H$_5$) without crossing too high kinetic barriers ($\sim$5 $k_{\rm B}T$, shown as black arrows), whereas the transition from M' to N involves several intermediates (H$_{ijk}$, H$_{ij}$ and H$_i$) and at least one high kinetic barrier ($\sim$10 $k_{\rm B}T$, shown as red arrows). Consequently, at an intermediate value of the pulling speed (not too far from equilibrium), unfolding events from a misfolded state that show rescue to N probably correspond to the unfolding of M''. To conclude, it is crucial to be out of equilibrium to favor misfolded states, but not too far so that rescue to N from M'' is preferential.

% \clearpage

If the distinction between M' and M'' is not made and a coarse grained state M is considered (M=M'$\cup$M''), a free energy for M that is nearly equal to that of M' is obtained. That is in accordance with the fact that the Boltzmann average of the free energies of M' and M'',
\begin{align}
\Delta G_M &=-k_{\rm B} T \log\left[ \exp{\left(-\frac{\Delta G_{M'}}{k_{\rm
  B}T}\right)}+\exp{\left(-\frac{\Delta G_{M''}}{k_{\rm B}T}\right)}\right]\nonumber\\
&\simeq \Delta G_{M'}= 61.97 k_{\rm B}T
\end{align}
is dominated by the lowest free energy among the two. In fact, $\Delta G_{M''}$
exceeds $\Delta
G_{M'}$ by $10$ $k_{\rm B}T$ units (Table 1, main text). 

% \clearpage

\section{Effect of neglecting the term $log(\phi_F^{A\rightarrow B}/\phi_R^{A\leftarrow B})$}

\begin{figure}[ht]
 \centering
\includegraphics[scale=1]{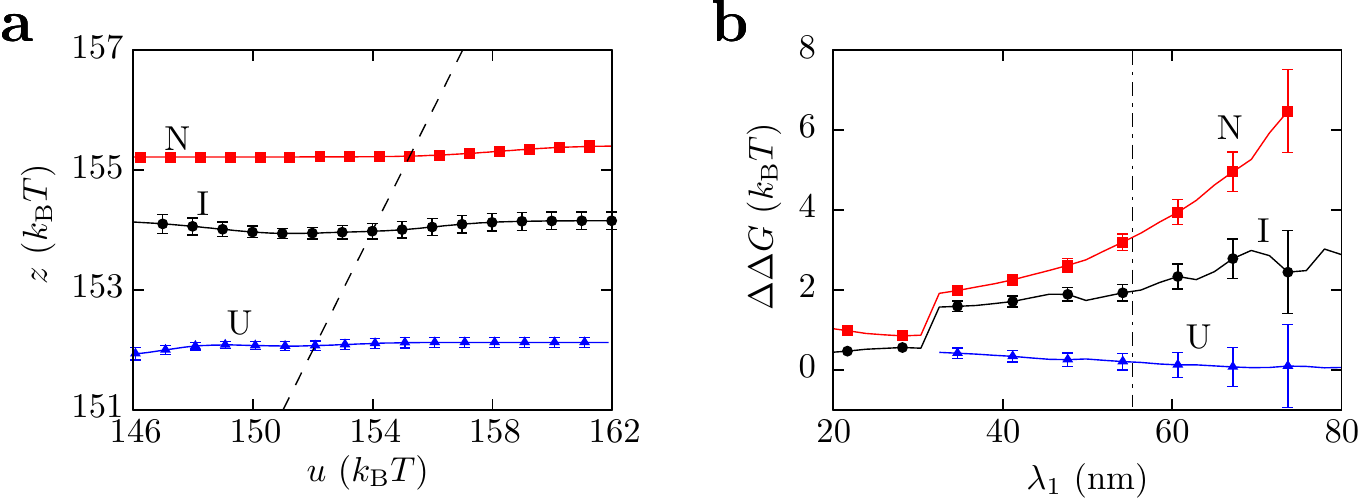}
\caption{\textbf{Molecule I1. Effect of neglecting the prefactor $\mathbf{\phi_F^{A\rightarrow B}/\phi_R^{A\leftarrow B}}$.}
\textbf{a.} The acceptance ratio method is applied to the work values measured between $\lambda_0=0$ to $\lambda_1=55.6$ nm. 
\textbf{b.} Difference between the free energy of each state, $\Delta G_{ {\rm N}B}$, $B=$ N (red), I (black), U (blue) and the free energy of the system $\Delta G$. Note the change with respect to the results shown in Fig. 2c and 2d (right panel) in the main text. Error bars were obtained from the bootstrap method.
}\label{fig: S3-n}
\end{figure}

The consequences of neglecting the correction $\log\left(\phi_F^{{\rm N}\to B}/\phi_R^{{\rm N}\leftarrow B}\right)$ can be seen in Fig. \ref{fig: S3-n}a, where results for the acceptance ratio method applied to work measurements between $\lambda_0=0$ and $\lambda_1=55.6$ nm are shown. Note that the relative stability of the three states N, I and U change as compared to Fig. 2c in the main paper. The same trend can be observed in Fig. \ref{fig: S3-n}b, where the free energy branches of the three kinetic states calculated by neglecting the correction are shown. These results should be compared with Fig. 2d (right inset) in the main paper. 

\clearpage

\section{Contributions to the evaluation of the free energy at zero force}

\begin{center}
 \begin{tabular}{|c|r@{$\pm$}lr@{$\pm$}lr@{$\pm$}l||r@{$\pm$}lr@{$\pm$}l|}
\hline
 & \multicolumn{6}{c||}{I1} & \multicolumn{4}{c|}{M1} \\
 & \multicolumn{6}{c||}{$\lambda_0$=0, $\lambda_1$=55.6 nm} & \multicolumn{4}{c|}{$\lambda_0$=0, $\lambda_1$=130.0 nm} \\
\cline{2-11}
 & \multicolumn{2}{c}{N$\rightarrow$N} & \multicolumn{2}{c}{N$\rightarrow$I} & \multicolumn{2}{c||}{N$\rightarrow$U} & \multicolumn{2}{c}{N$\rightarrow$U} & \multicolumn{2}{l|}{M$\rightarrow$U} \\
\hline
$\langle W\rangle_F^{A\rightarrow B}$ & 156 & 1 & 155 & 1 & 154 & 1 & 468 & 2 & 260 & 1\\
$\langle W\rangle_R^{A\leftarrow B}$  & 155 & 1 & 152 & 1 & 149 & 1 & 465 & 1 & 459 & 1\\
$\log\left(\phi_F^{A\rightarrow B}/\phi_R^{A\leftarrow B}\right)$ & \multicolumn{2}{c}{-0.11} & \multicolumn{2}{c}{-2.77} & \multicolumn{2}{c||}{-3.12} & \multicolumn{2}{c}{2.16} & \multicolumn{2}{l|}{\hspace{.06cm} 0.12} \\
$\Delta G_{AB}^{\rm EFR}$ & 156 & 1 & 157 & 1 & 156 & 1 & 464 & 2 & 459 & 1 \\
$\Delta G_{AB}^{\rm Crooks}$ & 156 & 1 & 154 & 1 & 152 & 1 & 466 & 2 & 460 & 1 \\
$\langle W_{\rm diss} \rangle_F^{A\rightarrow B}$ & 0 & 1 & 2 & 1 & 1 & 1 & 3 & 2 & 0 & 1 \\
$\langle W_{\rm diss} \rangle_R^{A\leftarrow B}$  & 0 & 1 & 5 & 1 & 6 & 1 & 1 & 2 & 0 & 1 \\
$W_{hb}$ & 156 & 3 & 115 & 3 & 83 & 3 & 381 & 3 & 392 & 3 \\
$W_{\rm ssDNA}$ & \multicolumn{2}{c}{0} & \multicolumn{2}{c}{9.3} & \multicolumn{2}{c||}{18.4} & \multicolumn{2}{c}{22.4} & \multicolumn{2}{l|}{\hspace{3mm}21.0 } \\
$\Delta G_{AB}^{0 ({\rm EFR})}$ & 0 & 3 & 30 & 3 & 55 & 3 & 61 & 3 & 49 & 3 \\
\hline
\end{tabular}
\end{center}
% % % % % % % % % % % % % % % % % % % % % % 
\begin{table}[ht]
\begin{tabular}{|c|r@{$\pm$}lr@{$\pm$}lr@{$\pm$}lr@{$\pm$}l||r@{$\pm$}lr@{$\pm$}lr@{$\pm$}l|}
\hline
 & \multicolumn{8}{c||}{I2} & \multicolumn{6}{c|}{M2} \\
 & \multicolumn{8}{c||}{$\lambda_0$=0, $\lambda_1$=148.0 nm} & \multicolumn{6}{c|}{$\lambda_0$=0, $\lambda_1$=230.0 nm} \\
\cline{2-15}
 & \multicolumn{2}{c}{N$\rightarrow$N} & \multicolumn{2}{c}{N$\rightarrow$I'} & \multicolumn{2}{c}{N$\rightarrow$I''} & \multicolumn{2}{c||}{N$\rightarrow$U} & \multicolumn{2}{c}{N$\rightarrow$U} & \multicolumn{2}{c}{M'$\rightarrow$U}  & \multicolumn{2}{l|}{M''$\rightarrow$U} \\
\hline
$\langle W\rangle_F^{A\rightarrow B}$ & 544 & 1 & 543& 1 & 541&1 & 544&1 & 623&1 & 583&1 & 611&1 \\
$\langle W\rangle_R^{A\leftarrow B}$  & 533&3 & 508&1 & 507&1 & 503&1 & 594&1 & 579&1 & 580&1 \\
$\log\left(\phi_F^{A\rightarrow B}/\phi_R^{A\leftarrow B}\right)$ & \multicolumn{2}{c}{-0.18} &  \multicolumn{2}{c}{-3.03} & \multicolumn{2}{c}{-2.42} & \multicolumn{2}{c||}{-3.72} & \multicolumn{2}{c}{0.71} & \multicolumn{2}{c}{1.75} & \multicolumn{2}{c|}{1.1} \\
$\Delta G_{AB}^{\rm EFR}$ & 543&3 & 527&1 & 526&1 & 528&1 & 612&1 & 579&1 & 590&1 \\
$\Delta G_{AB}^{\rm Crooks}$ & 544&3 & 525&1 & 525&1 & 526&1 & 613&1 & 581&1 & 592&1 \\
$\langle W_{\rm diss} \rangle_F^{A\rightarrow B}$ & 1&1 & 6&1 & 15&1 & 16&1 & 11&1 & 4&1 & 21&1 \\
$\langle W_{\rm diss} \rangle_R^{A\leftarrow B}$  & 10&3 & 19&1 & 21&1 & 25&1 & 18&1 & 0&1 & 10&1 \\
$W_{hb}$ & 543&4 & 469&6 & 417&6 & 365&6 & 480&2 & 487&3 & 485&3 \\
$W_{\rm ssDNA}$ & \multicolumn{2}{c}{0} & \multicolumn{2}{c}{18.9} & \multicolumn{2}{c}{30.5} & \multicolumn{2}{c||}{41.7} & \multicolumn{2}{c}{37.5} & \multicolumn{2}{c}{31.5} & \multicolumn{2}{c|}{35.5} \\
$\Delta G_{AB}^{0 ({\rm EFR})}$ & 0&4 & 40&6 & 80&7 & 125&7 & 95&3 & 60&3 & 69&3 \\
\hline
\end{tabular}
\caption{\textbf{Relevant energy contributions.}
For each molecule, we show the measured average forward and reverse works $\langle W\rangle_F^{A\rightarrow B}$ and $\langle W\rangle_R^{A\leftarrow B}$, the contribution to the free energy introduced by the correction term $\log\left(\phi_F^{A\rightarrow B}/\phi_R^{A\leftarrow B}\right)$, the free energy of the system obtained using the EFR $\Delta G_{AB}^{\rm EFR}$ (equation 1) and the Crooks fluctuation theorem (no correction term included) $\Delta G_{AB}^{\rm Crooks}$ \cite{Crooks2000}, the average forward and reversed dissipated works $\langle W_{\rm diss} \rangle_F^{A\rightarrow B}$ and $\langle W_{\rm diss} \rangle_R^{A\leftarrow B}$, the reversible work performed to stretch the handles and to displace the bead in the optical trap $W_{hb}$, and reversible work needed for the
stretching of the ssDNA and the orientation of the hairpin stem $W_{\rm ssDNA}$ (see Methods). Finally, we present the free energy estimation at zero force obtained using the EFR, $\Delta G_{AB}^{0 ({\rm EFR})}$.
Error bars contain statistical and systematic errors. All the magnitudes are given in $k_{\rm B}T$.
}\label{tab: relevantnumbers}
\end{table}

\clearpage

\section{Extended Jarzynski equality}

If we multiply equation 1 with the reverse partial work distribution $P_R^{N\leftarrow B}(-W)$ and integrate over the work we obtain the Extended Jarzynski equality (EJE):

\begin{equation}
 e^{-\frac{\Delta G_{AB}}{k_{\rm B}T}}=\frac{\phi_F^{A\rightarrow B}}{\phi_R^{A\leftarrow B}}\left\langle e^{-\frac{W}{k_{\rm B}T}}\right\rangle_F^{A\rightarrow B}
\end{equation}

The EJE allows to evaluate $\Delta G_{AB}$ using only the forward partial work distribution.

In the case of I2, the EJE was applied to recover the free energy branches of the four states N, I', I'' and U. $\lambda_0$ was fixed at a value where $A$=N for all trajectories, and therefore $\phi_R^{{\rm N}\leftarrow B}=1$, and $\lambda_1$ was varied between 100 and 200 nm. The free energy of each state at a given value of $\lambda_1$ can be evaluated using:

\begin{equation}
\Delta G_{{\rm N}B}=-k_{\rm B}T\log \phi_F^{{\rm N}\rightarrow B}-k_{\rm B}T\log\left\langle e^{-\frac{W}{k_{\rm B}T}}\right\rangle_F^{{\rm N}\rightarrow B}\label{eq: DG-EJE}
\end{equation}

Similarly to the Jarzynski equality \cite{Jarzynski1997}, the EJE is strongly biased \cite{PalassiniRitort2011}. Using pulling experiments where kinetic states are partially equilibrated at $\lambda_1$ (dark curves in Fig. 4a) we can evaluate the free energy difference $\Delta G_{{\rm N}B}$ using the EFR (equation (1)) and the EJE (equation \eqref{eq: DG-EJE}) for the four different states. The difference between the two magnitudes is an estimation of the bias (Table \ref{tab: bias}). 

\begin{table}[ht]
\centering
\begin{tabular}{lccc}
\hline
 $B$ & $\Delta G_{{\rm N}B}^{EJE}$ ($k_{\rm B}T$) & $\Delta G_{{\rm N}B}^{EFR}$ ($k_{\rm B}T$) & Bias ($k_{\rm B}T$) \\
\hline
N   & 543$\pm$3 & 543$\pm$3 &  0$\pm$4 \\
I'  & 545$\pm$1 & 527$\pm$1 & 18$\pm$2 \\
I'' & 539$\pm$1 & 526$\pm$1 & 13$\pm$2 \\
U   & 547$\pm$1 & 528$\pm$1 & 19$\pm$2 \\
\hline
\end{tabular}
\caption{\textbf{Estimation of bias for I2.} Free energy difference $\Delta G_{{\rm N}B}$ obtained from the EFR and the EJE, and estimation of the bias, for each state $B$=N, I', I'' and U.}\label{tab: bias}
\end{table}

To evaluate free energy branches we applied equation \eqref{eq: DG-EJE} to F processes obtained from standard pulling experiments (where no intermediates are observed at extreme values of $\lambda$) by fixing $\lambda_0$=0 nm and varying $\lambda_1$ between 100 and 200 nm. The estimation of the bias, taken independent of $\lambda_1$, is subtracted from the resulting free energies for each state.

% % % % % % % % % % % % % % % % % % % % % % % % % % % % % % % % % % % % % 

\end{document}